\newif\iflncs

\iflncs
\documentclass[orivec]{llncs} 
\else
\documentclass[12pt]{article}
\usepackage[left=3cm,right=3cm]{geometry}
\fi

\PassOptionsToPackage{svgnames,x11names,dvipsnames}{xcolor}
\usepackage{xcolor}
\usepackage{tikz}
\usetikzlibrary{patterns}

\usepackage{color}
\usepackage{array,tabularx}
\usepackage{paralist}
\usepackage{enumerate}
\usepackage[algoruled,vlined,english]{algorithm2e}
\usepackage[color=green!20]{todonotes}
\usepackage{url}
\usepackage{xspace}
\usepackage[draft]{fixme}
\usepackage{verbatim}
\usepackage{graphicx}
\usepackage{amssymb}
\usepackage{amsmath} 
\usepackage{listings}
\usepackage[breaklinks]{hyperref}

\usepackage{authblk} %

\iflncs
\spnewtheorem{obs}{Observation}{\bfseries}{\itshape}
\spnewtheorem{prop}{Proposition}{\bfseries}{\itshape}
\spnewtheorem{corol}{Corollary}{\bfseries}{\itshape}
\spnewtheorem{fact}{Fact}{\bfseries}{\itshape}
\else
\usepackage{amsthm}
\newtheorem{theorem}{THEOREM}[section]

\newtheorem{lemma}[theorem]{LEMMA}

\theoremstyle{definition}
\newtheorem{definition}[theorem]{Definition}

\newtheorem{example}[theorem]{EXAMPLE}

\newtheorem{obs}[theorem]{Observation}
\newtheorem{corol}[theorem]{Corollary}
\fi

\let\com=\newcommand
\com{\bthm}{\begin{theorem}}
\com{\ethm}{\end{theorem}}
\com{\bdfn}{\begin{definition}}
\com{\edfn}{\end{definition}}
\com{\blem}{\begin{lemma}}
\com{\elem}{\end{lemma}}
\com{\bcor}{\begin{corol}}
\com{\ecor}{\end{corol}}
\com{\bexm}{\begin{example}}
\com{\eexm}{\ensuremath{\hfill\Box}\end{example}}
\com{\bobs}{\begin{obs}}
\com{\eobs}{\end{obs}}
\com{\bprop}{\begin{proposition}}
\com{\eprop}{\end{proposition}}
\com{\be}{\begin{enumerate}}
\com{\ee}{\end{enumerate}}

\com{\bprf}{\begin{proof}}
\iflncs
\com{\eprf}{\qed\end{proof}}
\else
\com{\eprf}{\end{proof}}
\fi

\newcommand{\myref}[2]{\hyperref[#1]{\arabic{#1}$_{#2}$}}

\let\vect=\vec
\renewcommand{\vec}[1]{\mathbf{#1}}
\newcommand{\tuple}[1]{\langle #1 \rangle}
\newcommand{\cvz}[0]{\mathbf{0}}  %

\newcommand{\ints}{\ensuremath{\mathbb Z}\xspace}

\newcommand{\rats}{\ensuremath{\mathbb Q}\xspace}

\newcommand{\convhull}[0]{\mathrm{conv.hull}}
\newcommand{\cone}[0]{\mathrm{cone}}

\newcommand{\transitions}{\poly{Q}}
\newcommand{\distransitions}{\poly{R}}
\newcommand{\dpm}[0]{R}  %
\newcommand{\dct}{\poly{D}} %

\newcommand{\trcv}[2]{\ensuremath{\bigl(\begin{smallmatrix}{#1}\hfill\\{#2}\hfill\end{smallmatrix}\bigr)}}

\let\tr=\trcv

\newcommand{\poly}[1]{{\mathcal #1}}
\newcommand{\inthull}[1]{{#1}_I}
\newcommand{\intpoly}[1]{I({#1})}

\newcommand{\slc}[0]{\ensuremath{\mathit{SLC}}\xspace}

\newcommand{\lrf}[0]{\ensuremath{\mathit{LRF}}\xspace}
\newcommand{\mlrf}[0]{M{\ensuremath{\Phi}}RF\xspace}
\newcommand{\mlrfs}[0]{M{\ensuremath{\Phi}}RFs\xspace}

\newcommand{\lrfs}[0]{\ensuremath{\mathit{LRFs}}\xspace}

\newcommand{\trans}[0]{{\mbox{\tiny T}}}

\newcommand{\while}[0]{\ensuremath{\mathtt{while}\xspace}}
\newcommand{\wdo}[0]{\ensuremath{\mathtt{do}}\xspace}

\newcommand{\mlrfsym}[0]{\ensuremath{\tau}\xspace}

\newcommand{\diff}[1]{\ensuremath{\Delta #1}}

\newcommand{\ptime}[0]{\ensuremath{\mathtt{PTIME}}\xspace}
\newcommand{\coNP}[0]{\ensuremath{\mathtt{coNP}}\xspace}

\newcommand{\irankfinder}[0]{i\textsc{RankFinder}\xspace}

\newcommand{\proj}[2]{{\mathtt{proj}_{#1}{(#2)}}}
\newcommand{\ccone}[0]{{\mathtt{rec.cone}}}

\newcommand{\nonegfunc}[1]{\ensuremath{{#1}^{\#}}}

\newcommand{\bkwopsym}[0]{\ensuremath{F}}
\newcommand{\bkwop}[1]{\ensuremath{{\bkwopsym}(#1)}}
\newcommand{\bkwopiter}[2]{\ensuremath{{\bkwopsym}^{#2}(#1)}}

\newcommand{\highlight}[2]{\begingroup\setlength{\fboxsep}{0.5pt}\colorbox{#1}{#2}\endgroup\xspace}

\newcommand{\pre}[2]{{\mathtt{pre}^{#1}{(#2)}}}  
\newcommand{\dellrf}[1]{\ensuremath{\mathtt{RF}(#1)}}

\newcommand{\mapsym}[0]{\ensuremath{{\cal U}}}
\newcommand{\repl}[2]{{#1}^{\langle {#2}\rangle}}
\newcommand\spwedge[0]{\;\wedge\;}

\title{Multiphase-Linear Ranking Functions and their Relation to Recurrent Sets}

\iflncs
\author{
Amir M. Ben-Amram\inst{1} \and Jes\'us J. Dom\'enech\inst{2} \and Samir Genaim\inst{2}
} 
\institute{
School of Computer Science, The Tel-Aviv Academic College, Israel\\
\and
DSIC, Complutense University of Madrid (UCM),  Spain
}
\else
\author[1]{Amir M. Ben-Amram\thanks{amirben@mta.ac.il}}
\author[2]{Jes\'us J. Dom\'enech\thanks{jdomenec@ucm.es}}
\author[2]{Samir Genaim\thanks{samir.genaim@fdi.ucm.es}}
\affil[1]{The Academic College of Tel-Aviv Yaffo}
\affil[2]{Complutense University of Madrid}
\fi

\begin{document}

\maketitle

\begin{abstract}
Multiphase ranking functions (\mlrfs) are tuples $\tuple{f_1,\ldots,f_d}$ of linear functions
that are often used to prove termination of
loops in which the computation progresses through a number of
``phases".
Our work provides new insights regarding such functions for loops
described by a conjunction of linear constraints (Single-Path Constraint loops).
The decision problem \emph{existence of a \mlrf} asks to determine
whether a given \slc loop admits a \mlrf; and the corresponding
\emph{bounded} decision problem restricts the search to \mlrfs of
depth $d$, where the parameter $d$ is part of the input.
The algorithmic and complexity aspects of the bounded problem have been
completely settled in a recent work.
In this paper we make progress regarding the existence
problem, without a given depth bound.
In particular, we present an approach that reveals some important
insights into the structure of these functions. Interestingly, it relates the
problem of seeking \mlrfs to that of seeking recurrent sets (used to prove non-termination).
It also helps in identifying classes of loops
for which \mlrfs are sufficient.
Our research has led to a new representation for
single-path loops, the \emph{difference polyhedron} replacing the customary \emph{transition polyhedron}.
 This representation yields new insights on \mlrfs and SLC loops in general. For example, a result on bounded \slc loops becomes straightforward.
\end{abstract}

\section{Introduction}
\label{sec:intro}

Proving that a program will not go into an infinite loop is one of the
most fundamental tasks of program verification, and has been the
subject of voluminous research. Perhaps the best known, and often
used, technique for proving termination is that of
\emph{ranking functions}.
This consists of finding a function that maps program states into the
elements of a well-founded ordered set, such that it decreases when
applied to consecutive states.
This implies termination since infinite descent is impossible in a
well-founded order.

Unlike termination of programs in general, which is undecidable, the
algorithmic problems of detection (deciding the existence) or
generation (synthesis) of a ranking function can well be solvable,
given certain choices of the program representation, and the class of
ranking function.
There is a considerable amount of research in this direction, in which
different kinds of ranking functions for different kinds of program
representations were considered.
In some cases the algorithmic problems have been completely settled,
and efficient algorithms provided, while other cases remain open.
Besides proving termination, some classes of ranking functions also
serve to bound the length of the computation (an \emph{iteration
  bound}), useful in applications such as cost analysis and loop
optimization~\cite{Feautrier92.1,ADFG:2010,DBLP:journals/jar/AlbertAGP11,BrockschmidtE0F16}.

We focus on \emph{single-path linear-constraint loops} (\slc loops),
 where a state is described by the values of a finite set of
numerical variables, and the effect of a transition (one iteration of
the loop) is described by a conjunction of \emph{linear constraints}.
We consider the setting of integer-valued variables, as well as
rational-valued (or real-valued) variables. Here is an example of this
loop representation (formally defined in Section~\ref{sec:prelim}); primed variables
$x',y',\dots$ refer to the state following the transition.
\begin{equation}
\label{eq:loop-xyz}
\while~(x_1 \ge -x_3)~\wdo~x_1'=x_1+x_2,\; x_2'=x_2+x_3,\; x_3'=x_3-1
\end{equation}
Note that by $x_1'=x_1+x_2$ we mean an equation, not an assignment
statement. The description of a loop might involve linear
inequalities rather than equations, and consequently be
nondeterministic.
It is a standard procedure to compile sequential code (or approximate
it) into such representation using various techniques. We assume the
``constraint loop" to be given, and do not concern ourselves with the
orthogonal topic of extracting such loops from general programs.
This constraint representation may be extended to represent branching
in the loop body, a so-called \emph{multiple-path loop}; in the
current work we do not consider such loops. However, \slc loops are
important, e.g., in approaches that reduce a question about a whole
program to questions about simple
loops~\cite{Harrison1969book,leroux2005flat,CPR06,CGPRV2007,cookGLRS2008};
see~\cite{Ouaknine2015siglog} for references that show the importance
of such loops in other fields.

Several types of ranking functions have been suggested for \slc loops;
linear ranking functions (\lrfs) are probably the most known.  In this
case, we seek a function
$\rho(x_1,\dots,x_n) = a_1x_1+\dots+a_n x_n + a_0$ such that
\begin{inparaenum}[(i)]
\item\label{intro:lrf1} $\rho(\vec{x}) \ge 0$ for any valuation $\vec{x} = \tuple{x_1,\ldots,x_n}$ that satisfies
  the loop constraints (i.e.,  an enabled state); and
\item\label{intro:lrf2} $\rho(\vec{x})-\rho(\vec{x}') \ge 1$ for any transition leading from $\vec{x}$ to
  $\vec{x}' = \tuple{x_1',\ldots,x_n'}$.
\end{inparaenum}
The algorithmic problems of existence and synthesis of \lrfs has been
completely
settled~\cite{Feautrier92.1,DBLP:conf/pods/SohnG91,DBLP:conf/tacas/ColonS01,DBLP:conf/vmcai/PodelskiR04,Ben-AmramG13jv},
for both integer-valued and rational-valued variables, not only for
\slc loops but rather for control-flow graphs.

\lrfs do not suffice for all terminating \slc loops, e.g.,
Loop~\eqref{eq:loop-xyz} does not have a \lrf, and in such case, one
may resort to an argument that combines several linear functions to
capture a more complex behavior.
A common such argument is one that uses \emph{lexicographic ranking
  functions}, where a tuple of linear functions is required to
decreases lexicographically when moving from one state to another.
In this paper we are interested in a special case of the lexicographic
order argument that is called \emph{Multiphase ranking functions}
(\mlrf for short).
Intuitively, a \mlrf is a tuple $\tuple{f_1,\ldots,f_d}$ of linear
functions that define phases of the loop that are linearly ranked, as
follows:
$f_1$ decreases on all transitions, and when it becomes negative $f_2$
starts to decrease, and when $f_2$ becomes negative, $f_3$ starts to
decrease, etc.
For example, Loop~\eqref{eq:loop-xyz} has the \mlrf $\tuple{x_3+1,x_2+1,x_1}$.
The general definition (see Section~\ref{sec:prelim}) allows for an
arbitrary number $d$ of linear components; we refer to $d$ as
\emph{depth}, intuitively the number of phases.

The decision problem \emph{Existence of a \mlrf} asks to determine
whether a given \slc loop admits a \mlrf. The \emph{bounded} decision
problem restricts the search to \mlrfs of depth $d$, where the
parameter $d$ is part of the input.
The complexity and algorithmic aspects of the bounded version of the
\mlrf problem has been completely settled in~\cite{Ben-AmramG17}. 
The decision problem is \ptime for \slc loops with
rational-valued variables, and \coNP-complete for \slc loops with
integer-valued variables;  synthesizing \mlrfs, when they exist,
can be performed in polynomial and exponential time, respectively.
In addition, \cite{Ben-AmramG17} shows that for \slc loops \mlrfs has the same power
as general lexicographic orders, and that, surprisingly, \mlrfs induce
a linear bound on the number of iterations of \slc loops.
The problem of deciding whether a given \slc admits a \mlrf, without a
given bound on the depth, is still open.

In practice, termination analysis tools search for \mlrfs starting
by depth $1$ and incrementally increase the depth until they find
one, or reach a predefined limit, after which the returned answer is
\emph{don't know}.
Clearly, finding a theoretical upper-bound on the depth of a \mlrf,
given the loop, would also settle this problem. As shown
in~\cite{Ben-AmramG17}, such bound must depend not only on the number
of constraints or variables, but rather on the coefficients used in
the constraints.

In this paper we make progress towards solving the problem of
\emph{existence of a \mlrf}, i.e., seeking a \mlrf without a given
bound on the depth. In particular, we present an algorithm for seeking
\mlrfs that reveals important insights on the structure of these
ranking functions.
In a nutshell, the algorithm stars from the set of transitions of the
given \slc loop, which is a polyhedron, and iteratively removes
transitions $(\vec{x},\vec{x}')$ such that
$\rho(\vec{x})-\rho(\vec{x}')>0$ for some function
$\rho(\vec{x})=\vect{a}\cdot\vec{x}+b$ that is non-negative on all
enabled states.
The process continues iteratively, since
after removing some transitions, more
functions $\rho$ may satisfy the non-negativity condition,
and they may eliminate additional transitions in
the next iteration.
When all transitions are eliminated in a finite number of iterations, we can construct
a \mlrf using the $\rho$ functions; and when reaching a situation in
which no transition can be eliminated, we prove that we have actually
reached a recurrent set that witnesses nontermination.

The algorithm always finds a \mlrf if one exists, and in many cases it
finds a recurrent set (see the discussion on experiments in
Section~\ref{sec:conc}) when the loop is not terminating, however, it
is not a decision procedure as it diverges in some cases.
Nonetheless, our algorithm provides important insights on the
structure of \mlrfs. Apart from revealing a relation between seeking
\mlrfs and seeking recurrent sets of a particular form, these insights
are also useful for finding classes of \slc loops for which, when
terminating, there is always a \mlrf. We shall prove this for two
classes of \slc loops that have been considered before in previous work.

Our research, in addition, has led to a new representation for \slc
loops, that we refer to as the \emph{displacement} representation,
that provides us with new tools for studying termination of \slc loops
in general, and existence of a \mlrf in particular.
In this representation a transition $(\vec{x},\vec{x}')$ is
represented as $(\vec{x},\vec{y})$ where $\vec{y}=\vec{x}'-\vec{x}$.
We show that using this representation our algorithm can be formalized
in a simple way that avoids computing the $\rho$ functions mentioned
above (which might be expensive), and such that the existence of a
\mlrf is then equivalent to checking for unsatisfiability of a linear constraint system.
As an evidence on the usefulness of this representation in general, we
show that some nontrivial observations on termination of bounded \slc loops
are made straightforward in this representation, while they are not
easy to see in the normal representation.

The article is organized as follows.  Section~\ref{sec:prelim} gives
precise definitions and necessary background. Section~\ref{sec:alg}
describes our algorithm and its possible
outcomes. Section~\ref{sec:completeness} discusses classes of \slc
loops for which, when terminating, there is always a
\mlrf. Section~\ref{sec:dp} discusses the displacement representation
for \slc loops. Finally, in Section~\ref{sec:conc} we conclude and
discuss some related work.

\section{Preliminaries}
\label{sec:prelim}

In this section we give the fundamental definitions for this paper:
recall some definitions regarding (integer) polyhedra and mention some
important properties of these definitions, we define the class of
loops we study, the type of ranking functions, and the notion of
recurrent sets.

\subsection{Polyhedra}
\label{sec:prelim:poly}

A \emph{rational convex polyhedron} $\poly{P} \subseteq \rats^n$
(\emph{polyhedron} for short) is the set of solutions of a set of
inequalities $A\vec{x} \le \vec{b}$, namely
$\poly{P}=\{ \vec{x}\in\rats^n \mid A\vec x \le \vec b \}$, where
$A \in \rats^{m \times n}$ is a rational matrix of $n$ columns and $m$
rows, $\vec x\in\rats^n$ and $\vec b \in \rats^m$ are column vectors
of $n$ and $m$ rational values respectively.
We say that $\poly{P}$ is specified by $A\vec{x} \le \vec{b}$.  If
$\vec{b}=\cvz$, then $\poly{P}$ is a \emph{cone}.
The set of \emph{recession directions} of a polyhedron $\poly{P}$
specified by $A\vec{x} \le \vec b$, also known as its \emph{recession
  cone}, is the set $\ccone(\poly{P}) = \{ \vec{y}\in\rats^n \mid
A\vec{y} \le \vec{0}\}$.

For a given polyhedron $\poly{P} \subseteq \rats^n$ we let
$\intpoly{\poly{P}}$ be $\poly{P} \cap \ints^n$, i.e., the set of
integer points of $\poly{P}$. The \emph{integer hull} of $\poly{P}$,
commonly denoted by $\inthull{\poly{P}}$, is defined as the convex
hull of $\intpoly{\poly{P}}$, i.e., every rational point of
$\inthull{\poly{P}}$ is a convex combination of integer points.
It is known that $\inthull{\poly{P}}$ is also a polyhedron, and that
$\ccone(\poly{P})=\ccone(\inthull{\poly{P}})$.
An \emph{integer polyhedron} is a polyhedron $\poly{P}$ such that
$\poly{P} = \inthull{\poly{P}}$. We also say that such an integer
polyhedron is \emph{integral}.

Polyhedra also have a \emph{generator representation} in terms of
vertices and rays, written as
$\poly{P} = \convhull\{\vec x_1,\dots,\vec x_m\} + \cone\{\vec
y_1,\dots,\vec y_t\} \,.$
This means that $\vec x\in \poly{P}$ iff
$\vec x = \sum_{i=1}^m a_i\cdot \vec x_i + \sum_{j=1}^t b_j\cdot \vec
y_j$ for some rationals $a_i,b_j\ge 0$, where $\sum_{i=1}^m a_i = 1$.
Note that $\vec y_1,\dots,\vec y_t$ are the recession directions of
$\poly{P}$, i.e., $\vec{y}\in\ccone(\poly{P})$ iff
$\vec{y}=\sum_{j=1}^t b_j \cdot \vec{y}_j$ for some rationals
$b_j\ge 0$.
If $\poly{P}$ is integral, then there is a generator representation in
which all $\vec{x}_i$ and $\vec{y}_j$ are integers.

Let $\poly{P}\subseteq\rats^{n+m}$ be a polyhedron, and let
$\trcv{\vec{x}}{\vec{y}} \in \poly{P}$ be such that
$\vec{x}\in\rats^{n}$ and $\vec{y}\in\rats^{m}$.
The \emph{projection} of $\poly{P}$ onto the $\vec{x}$-space is
defined as
$\proj{\vec{x}}{\poly{P}}=\{\vec{x}\in\rats^n \mid \exists
\vec{y}\in\rats^{m} ~\mbox{such that}~ \trcv{\vec{x}}{\vec{y}} \in
\poly{P}\}$.
We will need the following lemmas later.

\blem 
\label{lem:proj.cone}
$\proj{\vec{x}}{\ccone(\poly{P})} = \ccone(\proj{\vec{x}}{\poly{P}}) $.
\elem

\bprf
A polyhedron $\poly{P}$ whose variables are split into two sets,
$\vec{x}$ and $\vec{y}$, can be represented in the form $A\vec{x} +
G\vec{y} \le \vec{b}$ for matrices $A$, $G$ and a vector $\vec b$ of
matching dimensions. 
Then~\cite[Theorem 11.11]{50yearsChap11} states that
$\proj{\vec{x}}{\poly{P}}$ is specified by the constraints $V(\vec{b}
- A\vec{x}) \ge \vec{0}$ for a certain matrix $V$ determined by $G$
only. From this it follows that $\ccone(\proj{\vec{x}}{\poly{P}}) = \{
\vec{x} \ : \ V\!A\vec{x} \le \vec{0} \}$.  But we can also apply the
theorem to $\ccone(\poly{P})$, which is specified by $A\vec{x} +
G\vec{y} \le \vec{0}$, and we get the same result
$\proj{\vec{x}}{\ccone(\poly{P})} = \{ \vec{x} \ : \ V\!A\vec{x} \le \vec{0}
\}$.
\eprf

\blem
\label{lem:posfuncstr}
Given a polyhedron $\poly{P}\ne\emptyset$, and linear functions $\rho_1,\ldots,\rho_k$ such that
\begin{enumerate}[\upshape(\itshape i\upshape)]
\item \label{posfuncstrAss1}
 $\vec{x}\in\poly{P} \rightarrow \rho_1(\vec{x}) > 0
  \vee\cdots\vee \rho_{k-1}(\vec{x}) > 0 \vee \rho_k(\vec{x}) \geq 0$
\item $\vec{x}\in\poly{P} \not\rightarrow \rho_1(\vec{x}) > 0 \vee\cdots\vee \rho_{k-1}(\vec{x}) > 0$
\end{enumerate}
There exist non-negative constants $\mu_1,\ldots,\mu_{k-1}$ such
that
$\vec{x}\in\poly{P} \rightarrow \mu_1 \rho_1(\vec{x})+\cdots+
 \mu_{k-1} \rho_{k-1}(\vec{x})+ \rho_{k}(\vec{x}) \geq 0$.
\elem

\bprf
Let $\poly{P}$ be $B\vec{x} \le \vec{c}$,
$\rho_i = \vect{a}_i\cdot\vec{x}-b_i$, then (\textit{i}) is equivalent to infeasibility of
\begin{equation}
  B\vec{x} \le \vec{c} \wedge A\vec{x} \le \vec{b} \wedge \vect{a}_k\cdot\vec{x} < b_k
\end{equation}
where $A$ consists of the $k-1$ rows $\vect{a}_i$, and $\vec{b}$ of corresponding $b_i$.
However, $B\vec{x}\le \vec{c} \land A\vec{x}\le \vec{b}$ is assumed to be feasible.

According to Motzkin's transposition theorem~\cite[Corollary 7.1k,
Page~94]{schrijver86}, this implies that there are row vectors
$\vect{\lambda}, \vect{\lambda}' \ge 0$ and a constant $\mu \ge 0$
such that the following is true:
\begin{equation}
\label{eq:Motz1}
\vect{\lambda}B+\vect{\lambda}'A + \mu a_k = 0 \land \vect{\lambda}\vec{c} + \vect{\lambda}'\vec{b} + \mu b_k \le 0 
 \land  ( \mu\ne 0 \lor  \vect{\lambda}\vec{c} + \vect{\lambda}'\vec{b} + \mu{b_k} < 0 )
\end{equation}
Now, if \eqref{eq:Motz1} is true, then for all $\vec{x}\in\poly{P}$,
\begin{align*}
(\sum_i \lambda'_i \rho_i (\vec{x}) ) + \mu \rho_k(\vec{x}) &= \vect{\lambda}'A\vec{x} - \vect{\lambda}'\vec{b}  + \mu a_k\vec{x} - \mu b_k \\
& = -\vect{\lambda}B\vec{x} - \vect{\lambda}'\vec{b} -\mu b_k  \ge  -\vect{\lambda}\vec{c} - \vect{\lambda}' \vec{b}  - \mu b_k \ge 0
\end{align*}
where if $\mu = 0$, the last inequality must be strict.  However, if
$\mu = 0$, then $\vect{\lambda}B + \vect{\lambda}'A = 0$, so by
feasibility of $B\vec{x}\le \vec{c}$ and $A\vec{x}\le \vec{b}$, 
this implies  %
 $\vect{\lambda}\vec{c} + \vect{\lambda}'\vec{b} \ge 0$, a
contradiction.  Thus, $(\sum_i \lambda'_i \rho_i ) + \mu \rho_k\ge 0$ on
$\poly{P}$ and $\mu> 0$. Dividing by $\mu$ we obtain the conclusion of
the lemma.
\eprf

\subsection{Single-Path Linear-Constraint Loops}
\label{sec:prelim:slcloops}

A \emph{single-path} linear-constraint loop (\slc loop) over $n$
variables $x_1,\ldots,x_n$ has the form
\begin{equation}
\label{eq:slc-loop}
  \mathit{while}~(B\vec{x} \le
  \vec{b})~\mathit{do}~ A\vec{x}+A'\vec{x}' \le \vec{c}\\
\end{equation}
where $\vec{x}=(x_1,\ldots,x_n)^\trans$ and
$\vec{x}'=(x_1',\ldots,x_n')^\trans$ are column vectors, and for some
$p,q>0$, $B \in {\rats}^{p\times n}$, $A,A'\in {\rats}^{q\times n}$,
$\vec{b}\in {\rats}^p$, $\vec{c}\in {\rats}^q$.
The constraint $B\vec{x} \le \vec{b}$ is called \emph{the loop
  condition} (a.k.a. the loop guard) and the other constraint is
called \emph{the update}.
The update is called \emph{deterministic} if, for a given $\vec x$
(satisfying the loop condition) there is at most one $\vec{x}'$
satisfying the update constraint. The update is called \emph{affine linear}
if it can be rewritten as
\begin{equation}
\label{eq:affine-update}
\vec{x}' = U\vec{x} + \vec{c}
\end{equation}
for a matrix $U \in \rats^{n\times n}$ and vector $\vec{c}\in\rats^n$.
We say that the loop is a \emph{rational loop} if $\vec{x}$ and
$\vec{x}'$ range over $\rats^n$, and that it is an \emph{integer loop}
if they range over $\ints^n$.  One could also allow variables to take
any real-number values, but for the problems we study,
where the constraints are expressed
by rational numbers, this very rarely differs from the rational case
(when it does, we comment on that explicitly).

We say that there is a transition from a state $\vec{x}\in\rats^n$ to
a state $\vec{x}'\in\rats^n$, if $\vec{x}$ satisfies the loop
condition and $\vec{x}$ and $\vec{x}'$ satisfy the update constraint.
A transition can be seen as a point
$\trcv{\vec{x}}{\vec{x}'} \in \rats^{2n}$, where its first $n$
components correspond to $\vec{x}$ and its last $n$ components to
$\vec{x}'$. For ease of notation, we denote $\tr{\vec{x}}{\vec{x}'}$
by $\vec{x}''$.
The set of all transitions $\vec{x}''\in \rats^{2n}$, of a given \slc
loop, will be denoted by $\transitions$ and is specified by the set of
inequalities $A'' \vec{x}'' \le \vec{c}''$ where
\begin{align*}
A''  & = \begin{pmatrix} B & 0 \\ A & A' \end{pmatrix}  &
\vec c'' & = \begin{pmatrix} \vec b \\ \vec c \end{pmatrix}
\end{align*}
and $B$, $A$, $A'$, $\vec{c}$ and $\vec{b}$ are those
of~\eqref{eq:slc-loop}.
We call $\transitions$ \emph{the transition polyhedron}.
For the purpose of this article, the loop is fully represented by this
polyhedron (in examples, we may use a more readable form as
\eqref{eq:slc-loop}).
For integer loops, the set of transitions is denoted by
$\intpoly{\transitions}$.

\subsection{Multi-Phase Ranking Functions}

An affine linear function $\rho: \rats^n \to \rats$ is a function of the
form $\rho(\vec{x}) = \vect{a}\cdot\vec{x} + b$ where
$\vect{a}\in\rats^n$ is a row vector and $b\in\rats$.
For a given function $\rho$, we define the function
$\diff{\rho}:\rats^{2n}\mapsto\rats$ as
$\diff{\rho}(\vec{x}'')=\rho(\vec{x})-\rho(\vec{x}')$.

\bdfn[\textnormal{\mlrf}]
\label{def:mlrf}
Given a set of transitions $T\subseteq\rats^{2n}$, we say that
$\mlrfsym=\tuple{\rho_1,\dots,\rho_d}$ is a \mlrf (of depth $d$) for $T$ if
for every $\vec{x}'' \in T$ there is an index $i$ such that:
\begin{alignat}{ 2 }
 \forall j \leq i \ .\   && \diff{\rho_j}(\vec{x}'') & \geq 1 \,, \label{eq:mlrf:1}\\[-0.5ex]
                      && \rho_i(\vec{x}) &\ge 0          \,, \label{eq:mlrf:2}\\[-0.5ex]
  \forall j < i \ .\                    && \rho_j(\vec{x}) &\le 0\,. \label{eq:mlrf:3}
\end{alignat}
We say that $\vec{x}''$ is \emph{ranked by} $\rho_i$ (for the minimal such $i$).
\edfn

It is not hard to see that a \mlrf $\tuple{\rho_1}$ of depth $d=1$ 
is a linear ranking function (\lrf).  On the other hand, if the \mlrf
is of depth $d>1$, it implies that if $\rho_1(\vec x) \ge 0$, transition
$\vec{x}''$ is ranked by $\rho_1$, while if $\rho_1(\vec x) < 0$,
$\tuple{\rho_2,\dots,\rho_d}$ becomes a \mlrf.  This agrees with the
intuitive notion of ``phases.''  We further note that, for loops
specified by polyhedra (namely, \slc loops as in~(\ref{eq:slc-loop})),
making the inequality \eqref{eq:mlrf:3} strict results in the same
class of ranking functions (we chose the definition that is easier to
work with), and, similarly, we can replace~\eqref{eq:mlrf:1} by
$\diff{\rho_j}(\vec{x}'') > 0$, obtaining an equivalent definition (up to
multiplication of the $\rho_i$ by some constants).
We say that $\mlrfsym$ is \emph{irredundant} if removing any component
invalidates the \mlrf.  Finally, it is convenient to allow an empty
tuple as a \mlrf, of depth $0$, for the empty set.

The decision problem \emph{Existence of a \mlrf} asks to determine
whether a given \slc loop admits a \mlrf.
The \emph{bounded} decision problem %
restricts the search to \mlrfs of depth at most $d$, where the
parameter $d$ is part of the input.

\subsection{Recurrent Sets}
\label{sec:prelim:recset}

A recurrent set is a set of states that witnesses nontermination of a
given \slc loop $\transitions$.
It is commonly defined as a set of states
$S \subseteq \proj{\vec{x}}{\transitions}$ where for any
$\vec{x}\in S$ there is $\vec{x}'\in S$ such that
$(\vec{x},\vec{x}') \in\transitions$.
This clearly proves the existence of an infinite run.
In this article we use a slightly different notion that uses
transitions instead of states.

\bdfn
\label{def:recset}
Give a \slc loop $\transitions$, we say that
$S \subseteq \transitions$ is a \emph{recurrent set of transitions} if
$\proj{\vec{x}'}{S} \subseteq \proj{\vec{x}}{S}$.
\edfn

Note that $\proj{\vec{x}}{S}$ is a recurrent set of states as
explained above.
It is known that the largest recurrent set (both for states and
transitions) of a given \slc loop is convex. In the rest of this
article we will mostly be interested in the case in which $S$ is a
closed polyhedron.

\section{An algorithm for inferring \mlrfs}
\label{sec:alg}

In this section we describe our algorithm for deciding the existence of
(and constructing) \mlrfs, which is also able to find recurrent sets
for certain nonterminating \slc loops.
In what follows we assume a given \slc loop $\transitions$ where
variables range over the rationals (or reals), the case of integer
variables is always discussed after considering the rational case.

Let us start with an intuitive description of the algorithm and its
possible outcomes.
Our work started with the following crucial observation: given linear
functions $\rho_1,\ldots,\rho_l$ such that
\begin{itemize}
\item $\rho_1,\ldots,\rho_l$ are nonnegative over
  $\proj{\vec{x}}{\transitions}$, i.e., over all enabled states;
\item for some $\rho_i$, we have $\diff{\rho_i}(\vec{x}'')>0$ for at least
  one transition $\vec{x}''\in\transitions$; and
\item
  $\transitions'=\transitions\wedge\diff{\rho_1}(\vec{x}'')\le
  0\wedge\cdots\wedge\diff{\rho_l}(\vec{x}'')\le 0$ has a \mlrf of depth
  $d$
\end{itemize}
then $\transitions$ has a \mlrf of depth at most $d+1$.  The proof of
this lemma is constructive, i.e., given a \mlrf $\tau'$ for
$\transitions'$, we can construct a \mlrf $\tau$ for $\transitions$
using conic combinations of the components of $\tau'$ and
$\rho_1,\ldots,\rho_l$.

Let us assume that we have a procedure $\bkwop{\transitions}$ that
picks some candidate functions $\rho_1,\ldots,\rho_l$, i.e., nonnegative
over $\proj{\vec{x}}{\transitions}$, and computes
$\bkwop{\transitions}=\transitions\wedge\diff{\rho_1}(\vec{x}'')\le
0\wedge\cdots\wedge\diff{\rho_l}(\vec{x}'')\le 0$.
Clearly, if $\bkwopiter{\transitions}{d}=\emptyset$, for some $d>0$,
then using the above observation we can conclude that $\transitions$
has a \mlrf of depth at most $d$.
Obviously, the difficult part in defining $\bkwopsym$ is how to find
functions $\rho_1,\ldots,\rho_l$, and, moreover, how to choose ones to
find the optimal depth $d$, equivalently, so that if $\transitions$ has a \mlrf of
optimal depth $d$ then 
$\bkwopiter{\transitions}{d}=\emptyset$.
For this, we note that the set of all nonnegative functions over
$\proj{\vec{x}}{\transitions}$ is a polyhedral cone, and thus it has
generators $\rho_1,\ldots,\rho_l$ that can be effectively
computed. These $\rho_1,\ldots,\rho_l$ turn to be the right candidates
to use. In addition, when using these candidates, we prove that for
cases in which we make no progress, i.e.,
we get  $\bkwopiter{\transitions}{i-1}=\bkwopiter{\transitions}{i}$, then we
have actually reached a recurrent set that witnesses nontermination.

The rest of this section is organized as follow. In
Section~\ref{sec:alg:mlrf} we present the algorithm and discuss how it
is used to decide existence of \mlrfs; in Section~\ref{sec:alg:recset}
we discuss how the algorithm can infer recurrent sets; and in
Section~\ref{sec:alg:limit} we discuss cases where the algorithm does
not terminate and raise some questions on what happens in the limit.

\subsection{Deciding existence of \mlrfs}
\label{sec:alg:mlrf}

We start by formally defining the set of all nonnegative functions
over a given polyhedron $\poly{S} \subseteq \rats^n$.

\bdfn
\label{def:nonneg}
The set of all nonnegative functions over a polyhedron $\poly{S} \subseteq \rats^n$, is defined as
$\nonegfunc{\poly{S}}= \{ (\vect{a},b) \in \rats^{n+1} \mid \forall\vec{x}\in \poly{S}.~\vect{a}\cdot\vec{x}+b \ge 0
\}.$
\edfn

It is known that $\nonegfunc{\poly{S}}$ is a polyhedral cone~\cite[p.~112]{schrijver86}.
Equivalently, it is generated by a finite set of rays
$(\vect{a}_1,b_1),\ldots,(\vect{a}_l,b_l)$.
The cone generated by $\vect{a}_1,\ldots,\vect{a}_l$ is known as
the dual of the cone $\ccone(\proj{\vec{x}}{\poly{S}})$ -- we make use
of this in Section~\ref{sec:dp}. These rays are actually the ones that
are important for the algorithm, as can be seen in the definition
below, however, in the definition of $\nonegfunc{\poly{S}}$ we included the $b_i$'s as they makes
some statements smoother.
Note that $\poly{S}$ is equal to the intersection of all
half-spaces defined the elements of $\nonegfunc{\poly{S}}$, i.e.,
$\poly{S}=\wedge\{ \vect{a}\cdot\vec{x}+b \ge 0 \mid (\vect{a},b)\in
\nonegfunc{\poly{S}}\}$, since it is a closed convex set.

\bdfn
\label{def:op}
Let $\transitions$ be a \slc loop, and define
$$\bkwop{\transitions} = \transitions\spwedge\vect{a}_1\cdot\vec{x}-\vect{a}_1\cdot\vec{x}'\le0\spwedge\cdots\spwedge\vect{a}_l\cdot\vec{x}-\vect{a}_l\cdot\vec{x}'\le 0
$$
where $(\vect{a}_1,b_1),\ldots,(\vect{a}_l,b_l)$ are the generators of
$\nonegfunc{\proj{\vec{x}}{\transitions}}$.
\edfn

Note that each $\vect{a}_i\cdot\vec{x}-\vect{a}_i\cdot\vec{x}'\le0$
above is actually $\diff{\rho_i}(\vec{x}'')\le0$ where
$\rho_i=\vect{a}_i\cdot\vec{x}+b_i\le0$.
Intuitively, $\bkwop{\transitions}$ removes from $\transitions$ all
transitions $\vec{x}''$ for which there is
$(\vect{a},b) \in \nonegfunc{\proj{\vec{x}}{\transitions}}$ such that
$\vect{a}\cdot\vec{x}-\vect{a}\cdot\vec{x}' > 0$.
This is because any $(\vect{a},b) \in
\nonegfunc{\proj{\vec{x}}{\transitions}}$ is a conic combination of
$(\vect{a}_1,b_1),\ldots,(\vect{a}_l,b_l)$, and thus for some $i$ we
must have $\vect{a}_i\cdot\vec{x}-\vect{a}_i\cdot\vec{x}'> 0$,
otherwise we would have $\vect{a}\cdot\vec{x}-\vect{a}\cdot\vec{x}'=0$.

\bexm
\label{ex:oper}
Consider the \slc loop~\eqref{eq:loop-xyz},
whose transition polyhedron is
$\transitions = \{ x_1 \ge -x_3, x_1'= x_1 + x_2,\ x_2'=x_2+x_3,\ x_3=x_3'-1\}$.
The generators of $\nonegfunc{\proj{\vec{x}}{\transitions}}$ are
$\{(1,0,1,0),(0,0,0,1)\}$.
That is, the corresponding nonnegative functions are
$\rho_1(x_1,x_2,x_3)=x_1+x_3$ and $\rho_2(x_1,x_2,x_3)=1$ (recall
that, in each generator, the last component corresponds to the free
constant $b$, and the rest to $\vect{a}$). Computing
$\bkwop{\transitions}$ results in:
\begin{equation}
\transitions' = \transitions \wedge \diff{\rho_1}(\vec{x}'') \le 0 \wedge \diff{\rho_2}(\vec{x}'') \le 0
              = \transitions \wedge (x_1+x_3)-(x_1'+x_3') \le 0
\end{equation}
This eliminates any transition for which the quantity $x_1+x_3$
decreases.  \eexm

In what follows we aim at showing that $\transitions$ has a \mlrf of
optimal depth $d$ iff $\bkwopiter{\transitions}{d}=\emptyset$. We
first state some auxiliary lemmas.

\blem
\label{lem:mlrf:1}
If $\transitions'=\bkwop{\transitions}$ has a \mlrf of depth at most $d$, then
$\transitions$ has a \mlrf of depth at most $d+1$.
\elem

\bprf
Consider the generators
$(\vect{a}_1,b_1),\ldots,(\vect{a}_l,b_l)$ used in
Definition~\ref{def:op}, and let
$\rho_i(\vec{x})=\vect{a}_i\cdot\vec{x}+b_i$.
We have
$\transitions'=\transitions\wedge \diff{\rho_1}(\vec{x}'')\le 0 \wedge
\cdots \wedge \diff{\rho_l}(\vec{x}'')\le 0$. We assume that no
$\rho_i$ is redundant, otherwise we take an irredundant subset.
Note that simply appending $\rho_1,\ldots,\rho_l$ to a \mlrf $\tau$ of
$\transitions'$ does not always produce a \mlrf, since the components
of $\tau$ are not guaranteed to decrease over
$\transitions\setminus\transitions'$.
Let $\tau=\tuple{g_1,\ldots,g_d}$ be a \mlrf for $\transitions'$, and
w.l.o.g.  assume that it is of optimal depth, we show how to construct
a \mlrf $\tuple{g_1'+1,\ldots,g_d'+1,g_{d+1}}$ for $\transitions$. We
construct the components one by one.

If $g_1$ is decreasing over $\transitions$, we define
$g_1'(\vec{x})=g_1(\vec{x})$, otherwise we have
\begin{align}
\vec{x}''\in \transitions  \rightarrow & \diff{\rho}_1(\vec{x}'')>0 \vee \cdots \vee \diff{\rho}_l(\vec{x}'')>0
\vee \diff{g_1}(\vec{x}'') - 1 \geq 0 \\
\vec{x}''\in \transitions \not\rightarrow & \diff{\rho}_1(\vec{x}'')>0
\vee \cdots \vee \diff{\rho}_l(\vec{x}'')>0
\end{align}
and by Lemma~\ref{lem:posfuncstr} there are non-negative constants
$\mu_1,\ldots,\mu_l$ such that
\begin{equation}
  \vec{x}''\in \transitions \rightarrow \diff{g_1}(\vec{x}'') - 1 + \sum_{i=1}^{l} \mu_i\diff{\rho_i}(\vec{x''}) \ge 0.
\end{equation}
Define $g_1'(\vec{x})=g_1(\vec{x})+\sum_{i=1}^l \mu_i \rho_i(\vec{x})$.
Clearly we have
$\vec{x}''\in \transitions \rightarrow \diff{g_1'}(\vec{x}'') \geq
1$. Note also that, since $\rho_1,\ldots,\rho_l$ are non-negative on all
enabled states, $g_1'$ is non-negative on the states on which $g_1$ is
non-negative.
If $d>1$, we proceed with
\begin{equation}
\transitions^{(1)} = \transitions \cap \{\vec{x}''\mid g_1'(\vec{x})\le (-1)\}.
\end{equation}
If $g_2$ is decreasing over $\transitions^{(1)}$, let $g_2'=g_2$,
otherwise, since transitions in $\transitions'\cap\transitions^{(1)}$ are
ranked by $\tuple{g_2,\dots,g_d}$ we have
\begin{align}
\vec{x}''\in \transitions^{(1)} & \rightarrow \diff{\rho}_1(\vec{x}'')>0 \vee \cdots \vee \diff{\rho}_l(\vec{x}'')>0 \vee \diff{g_2}(\vec{x}'') - 1 \geq 0 \\
\vec{x}''\in \transitions^{(1)} & \not\rightarrow \diff{\rho}_1(\vec{x}'')>0
\vee \cdots \vee \diff{\rho}_l(\vec{x}'')>0
\end{align}
and again by Lemma~\ref{lem:posfuncstr} we can construct the desired
$g_2'$ as we did for $g_1'$.
In general, for any $j\le d$ we construct $g_{j+1}'$ such that
$\diff{g_{j+1}'}(\vec{x}'') \ge 1$ over
\begin{equation}
\transitions^{(j)} = \transitions  \cap \{\vec{x}''\in\rats^{2n}\mid g_1'(\vec{x}) \le (-1) \wedge \dots \wedge g_{j}'(\vec{x}) \le (-1) \}
\end{equation}
and $\vec{x}''\in \transitions \land g_j(\vec{x}) \ge 0 \rightarrow g_j'(\vec{x}) \ge 0$.
Finally we define
\begin{equation}
  \transitions^{(d)} = \transitions  \cap \{\vec{x}''\in\rats^{2n}\mid g_1'(\vec{x}) \le (-1) \wedge \cdots \wedge g_d'(\vec{x}) \le (-1) \}
\end{equation}
and note that
\begin{equation}
\label{prf:1:1}
  \vec{x}''\in \transitions^{(d)} \rightarrow \diff{\rho}_1(\vec{x}'')>0 \vee \cdots \vee \diff{\rho}_l(\vec{x}'')>0
\end{equation}
We assume that no $\rho_i$ is redundant in~\eqref{prf:1:1}, otherwise
we take an irredundant subset. Now from~\eqref{prf:1:1} we get
\begin{equation}
\vec{x}''\in (\transitions^{(d)} \wedge \diff{\rho}_1(\vec{x}'') \le
0\wedge\cdots\wedge\diff{\rho}_{l-1}(\vec{x}'') \le 0) \rightarrow
\diff{\rho}_l(\vec{x}'')>0
\end{equation}
and since the left-hand side is a polyhedron, there must be a constant $c>0$ such
that
\begin{equation}
\vec{x}''\in (\transitions^{(d)} \wedge \diff{\rho}_1(\vec{x}'') \le
0\wedge\cdots\wedge\diff{\rho}_{l-1}(\vec{x}'') \le 0) \rightarrow
\diff{\rho}_l(\vec{x}'')\ge c.
\end{equation}
W.l.o.g. we may assume that $c \ge 1$, otherwise we divide $\rho_l$ by $c$.
Then we have
\begin{align}
\vec{x}''\in \transitions^{(d)} \rightarrow & \diff{\rho}_1(\vec{x}'')>0 \vee \cdots \vee \diff{\rho}_{l-1}(\vec{x}'')>0 \vee \diff{\rho}_l(\vec{x}'') - 1 \ge 0\\
\vec{x}''\in \transitions^{(d)} \not\rightarrow & \diff{\rho}_1(\vec{x}'')>0 \vee \cdots \vee \diff{\rho}_{l-1}(\vec{x}'')>0 
\end{align}
and again by Lemma~\ref{lem:posfuncstr} we can construct
$g_{d+1} = \rho_l+\sum_{i=1}^{l-1}\mu_i\rho_i$ such that
$\vec{x}''\in \transitions^{(d)} \rightarrow
\diff{g_{d+1}}(\vec{x}'')\ge 1$. Moreover, $g_{d+1}$ is non-negative
over $\transitions^{(d)}$ and thus it ranks all transitions in
$ \transitions^{(d)}$.
Now, by construction, $\tau'=\tuple{g_1'+1,\dots,g_d'+1,g_{d+1}}$ is a \mlrf for
$\transitions$.
\eprf

\blem
\label{lem:mlrf:2}
If $\transitions$ has a \mlrf of depth $d$ then
$\transitions'=\bkwop{\transitions}$ has a \mlrf of depth at most
$d-1$.
\elem

\bprf
Let $\tau=\tuple{\rho_1,\ldots,\rho_k}$ be an \mlrf
for $\transitions$, of optimal depth $k\le d$. Without lose of generality we may assume that
$\rho_k$ is non-negative on all $\proj{\vec{x}}{\transitions}$, this
follows immediately from the definition of nested
\mlrf~\cite{LeikeHeizmann15}, which is a special case of \mlrf in
which the last component is nonnegative, and the fact that for the
case of \slc loops existence of a \mlrf implies the existence of a
nested \mlrf~\cite{Ben-AmramG17} of the same optimal depth.
Clearly $\tau'=\tuple{\rho_1,\ldots,\rho_{k-1}}$ is a \mlrf for
$\transitions\wedge \diff{\rho_k}(\vec{x}'')\le 0$ since transitions
that are ranked by $\rho_k$ are eliminated.
Now since $\rho_k$ is a conic combination of the generators of
$\nonegfunc{\proj{\vec{x}}{\transitions}}$ we have
$\transitions'=\bkwop{\transitions} \subseteq \transitions\wedge
\diff{\rho_k}(\vec{x}'')\le 0$ and thus $\tau'$ is a \mlrf for
$\transitions'$ as well.
\eprf

\blem
\label{lem:mlrf}
$\transitions$ has a \mlrf of depth $d$ iff
$\bkwopiter{\transitions}{d}=\emptyset$.
\elem

\bprf
For the first direction, suppose that $\transitions$ has a \mlrf of
depth at most $d$, then applying Lemma~\ref{lem:mlrf:2} iteratively we
must reach $\bkwopiter{\transitions}{k}=\emptyset$ for some $k \le d$,
thus $\bkwopiter{\transitions}{d}=\emptyset$.
For the other direction, suppose
$\bkwopiter{\transitions}{d}=\emptyset$, then using
Lemma~\ref{lem:mlrf:1} we can construct a \mlrf of depth $d$.
It is easy to see that if $\bkwopiter{\transitions}{d}=\emptyset$ and
$\bkwopiter{\transitions}{d-1}\neq\emptyset$, then $d$ is the optimal
depth.
\eprf

\begin{algorithm}[t]
\caption{Deciding existence of \mlrfs and inferring recurrent sets}
\label{alg:mlrf}
\DontPrintSemicolon
\SetKwFunction{procdecidemlrf}{FindMLRF}
\BlankLine 
\procdecidemlrf{${\poly{\transitions}}$}\;
\Begin{
  \nl \lIf{\rm{(}$\transitions$ is empty\rm{)}}{\Return $\emptyset$}\label{alg:empty}
  \nl \Else{
    \nl Compute the generators $(\vect{a}_1,b_1),\ldots,(\vect{a}_l,b_l)$ of $\nonegfunc{\proj{\vec{x}}{\transitions}}$\label{alg:op:1}\\
    \nl $\transitions'$=$\transitions\spwedge\vect{a}_1\cdot\vec{x}-\vect{a}_1\cdot\vec{x}'\le 0\spwedge\cdots\spwedge\vect{a}_l\cdot\vec{x}-\vect{a}_l\cdot\vec{x}'\le 0$\label{alg:op:2}\\
    \nl \lIf{\rm{(}$\transitions'$ {=}{=} $\transitions$\rm{)}}{\Return $\transitions$}\label{alg:rs}
    \nl \lElse{\Return \procdecidemlrf($\transitions'$)}
  }
}
\end{algorithm}

Procedure~$\procdecidemlrf(\transitions)$ of Algorithm~\ref{alg:mlrf}
implements the above idea, it basically applies $\bkwopsym$ (lines
\ref{alg:op:1}-\ref{alg:op:2}) iteratively until it either reaches an
empty set (Line~\ref{alg:empty}) or stabilizes (Line~\ref{alg:rs}).
If it returns $\emptyset$ then $\transitions$ has a \mlrf and we can
construct one simply by invoking the polynomial-time procedure for
synthesizing nested \mlrfs as described in~\cite{Ben-AmramG17}, or
construct one as in the proof of Lemma~\ref{lem:mlrf:1}.
Note that, by Lemma~\ref{lem:mlrf}, if we bound the recursion depth by
a parameter $d$, then the algorithm is actually a decision procedure
for the existence of \mlrfs of at most depth $d$.
The case in which it returns a nonempty set is discussed in
Section~\ref{sec:alg:recset}.
Note that the complexity of Algorithm~\ref{alg:mlrf} is exponential
since computing the generators at Line~\ref{alg:op:1} might take
exponential time. In Section~\ref{sec:dp} we provide a polynomial-time
implementation that does not require computing the generators. 

For implementation purposes, we note that computing the generators of
$\nonegfunc{\proj{\vec{x}}{\transitions}}$ at Line~\ref{alg:op:1} can
be done without computing $\proj{\vec{x}}{\transitions}$ as follows.
Assume that $\transitions\equiv[A''\vec{x}\le\vec{c}'']$, using
Farkas' lemma we get that
$(\vect{a},b)\in\nonegfunc{\proj{\vec{x}}{\transitions}}$ iff it is
satisfies the constraint set
$\poly{C}\equiv[\vect{\lambda}A'' = (-\vect{a},\vect{0}) \wedge
\vect{\lambda}\vec{c} \le b\wedge \vect{\lambda}I \ge \vect{0}]$,
for some $\vect{\lambda}$ ($I$ is an identity matrix of appropriate dimension). $\poly{C}$ defines a cone
over the coordinates $\vect{\lambda},\vect{a}$ and $b$. We can compute its
generators using a standard algorithm, and then take the coordinates that correspond to
$(\vect{a},b)$.

\bexm
\label{ex:loop-xyz}
Let us apply the algorithm to Loop~\eqref{eq:loop-xyz}.  We start by
calling \procdecidemlrf with the  transition polyhedron
$\transitions=\{x_1\ge-x_3, x_1'=x_1+x_2, x_2'=x_2+x_3, x_3'=x_3-1\}$
and proceed as follows ($\transitions_i$ represents the polyhedron
passed in the $i$-th call to \procdecidemlrf):
\begin{small}
\[
  \begin{array}{|l|l|}
  \hline
  \multicolumn{1}{|c|}{\transitions_i}  & \multicolumn{1}{c|}{\mbox{Generators of}~\nonegfunc{\proj{\vec{x}}{\transitions_i}}} \\
\hline
    \transitions_0 {=} \transitions &
                     \{\mathbf{(1,0,1,0)},(0,0,0,1)\} \\
    \hline
    \transitions_1{=} \transitions_0\wedge(x_1+x_3)-(x_1'+x_3') \le 0 & 
        \{\mathbf{(0,1,0,-1)},(1,0,1,0),(0,0,0,1)\} \\
    \hline
    \transitions_2{=} \transitions_1\wedge x_2-x_2'\le 0 &
               \{\mathbf{(0,0,1,0)}, (0,1,0,-1),(1,0,1,0),(0,0,0,1)\} \\
    \hline
    \transitions_3{=}\transitions_2\wedge x_3-x_3'\le 0 = \emptyset& \\
  \hline
  \end{array}
\]
\end{small}%
Let us explain the above steps:

\begin{itemize}

\item $\transitions_0$ is not empty, so we compute the generators of
  $\nonegfunc{\proj{\vec{x}}{\transitions_0}}$, which define
  the nonnegative functions $\rho_1(x_1,x_2,x_3)=x_1+x_3$ and
  $\rho_2(x_1,x_2,x_3)=1$, and then compute
  $\transitions_1=\transitions_0\wedge\diff{\rho_1}(\vec{x}'') \le 0
  \wedge \diff{\rho_2}(\vec{x}'') \le 0$; and since it differs from
  $\transitions_0$ we recursively call
  $\procdecidemlrf(\transitions_1)$.
\item $\transitions_1$ is not empty, so we compute the generators of
  $\nonegfunc{\proj{\vec{x}}{\transitions_1}}$, which define the
  nonnegative function $\rho_3(x_1,x_2,x_3)=x_2-1$, and then compute
  $\transitions_2=\transitions_1\wedge \diff{\rho_3}(\vec{x}'')\le 0$;
  and since it differs from $\transitions_1$ we recursively call
  $\procdecidemlrf(\transitions_2)$. Note that the only new generator
  wrt.\ the previous iteration is the one in bold font, the others can
  be ignored since they have been used already when computing
  $\transitions_1$.
\item $\transitions_2$ is not empty, so we compute the generators of
  $\nonegfunc{\proj{\vec{x}}{\transitions_2}}$, which define the
  nonnegative function $\rho_4(x_1,x_2,x_3)=x_3$, and then compute
  $\transitions_3=\transitions_2\wedge\diff{\rho_4}(\vec{x}'')\le 0$; and since it
  differs from $\transitions_2$ we recursively call
  $\procdecidemlrf(\transitions_3)$.
\item $\transitions_3$ is empty, so we return $\emptyset$.
\end{itemize}
Since we have reached an empty set in $3$ iterations, we conclude that
Loop~\eqref{eq:loop-xyz} has a \mlrf of optimal depth $3$, e.g.,
$\tuple{x_3+1,x_2+1,x_1+x_3+1}$.
\eexm

Let us discuss now the case in which the variables range over the
integers, i.e., the set of integer transitions $\intpoly{\transitions}$.
It is know that $\intpoly{\transitions}$ has a \mlrf iff the integer
hull $\inthull{\transitions}$ of $\transitions$ has a \mlrf (over the
rationals)~\cite[Sect.~5]{Ben-AmramG17}. This leads to the following lemma.

\blem
\label{lem:mlrf:ints}
$\intpoly{\transitions}$ has a \mlrf of depth $d$ iff
$\bkwopiter{\inthull{\transitions}}{d}=\emptyset$.
\elem

\subsection{Inference of recurrent sets}
\label{sec:alg:recset}

Next we discuss the case in which $\procdecidemlrf(\transitions)$
returns a nonempty set of transition $\poly{S} \subseteq \transitions$
(Line~\ref{alg:rs}), and show that $\poly{S}$ is always a recurrent
set, implying that $\transitions$ is nonterminating. In
Section~\ref{sec:conc} we discuss an experimental evaluation regards
the use of Algorithm~\ref{alg:mlrf} for proving nontermination.

\blem
\label{lem:fixpoint}
Let $\poly{S}\subseteq \rats^{2n}$ be a polyhedron, if
$\poly{S}=\bkwop{\poly{S}}$ then $\poly{S}$ is a recurrent set.
\elem

\bprf
According Definition~\ref{def:recset}, we need to show that
$\proj{\vec{x}'}{\poly{S}} \subseteq \proj{\vec{x}}{\poly{S}}$.
Recall that since $\proj{\vec{x}}{\poly{S}}$ and
$\proj{\vec{x}'}{\poly{S}}$ are closed convex sets, each is an
intersection of half-spaces that are defined by the corresponding sets
$\nonegfunc{\proj{\vec{x}}{\poly{S}}}$ and
$\nonegfunc{\proj{\vec{x}'}{\poly{S}}}$, e.g.,
$\proj{\vec{x}}{\poly{S}} = \wedge\{ \vect{a}\cdot\vec{x}+b \ge 0 \mid
(\vect{a},b)\in \nonegfunc{\proj{\vec{x}}{\poly{S}}}\}$.
Thus, it is enough to show that
$\nonegfunc{\proj{\vec{x}}{\poly{S}}} \subseteq
\nonegfunc{\proj{\vec{x}'}{\poly{S}}}$.

Let $(\vect{a},b) \in \nonegfunc{\proj{\vec{x}}{\poly{S}}}$, we show that
$(\vect{a},b) \in \nonegfunc{\proj{\vec{x}'}{\poly{S}}}$ as well.
Define $\rho(\vec x) = \vect{a}\cdot{\vec x} + b$, which is
nonnegative over $\proj{\vec{x}}{\poly{S}}$, and note that, by
definition of $\bkwopsym$, since $\poly{S}=\bkwop{\poly{S}}$ we have
\[
\vec{x}''=(\vec{x},\vec{x'})\in \poly{S} \models 0 \le \rho(\vec{x}) \le \rho(\vec{x}')
\]
so $\vect{a}\cdot\vec{x}'+b \ge 0$ holds for any
$\vec{x}'\in \proj{\vec{x}'}{\poly{S}}$ and thus
$(\vect{a},b) \in \proj{\vec{x}'}{\poly{S}}$.
\eprf

\bcor
\label{cor:recurrent-set}
If $\procdecidemlrf(\transitions)$ returns $\poly{S} \neq \emptyset$
then $\poly{S}$ is a recurrent set, and thus $\transitions$ is
nonterminating.
\ecor

\bprf
This follows from the last lemma, since the algorithm returns a nonempty set
$\poly{S} \subseteq\transitions$ iff it finds one such that
$\poly{S}=\bkwop{\poly{S}}$ (Line~\ref{alg:rs} of \procdecidemlrf).
\eprf

\bexm
Let us apply the algorithm to the following loop, from~\cite{Tiwari:04}:
\begin{equation}
\label{eq:loop:tiwari}
\verb/while / (x_1-x_2 \ge 1) \verb/ do / x_1'= -x_1 + x_2,\  x_2'=x_2
\end{equation}
This loop does not terminate,
e.g., for $x_1=-1, x_2=-2$.
We call \procdecidemlrf with
$\transitions=\{x_1-x_2\ge1, x_1'=-x_1+x_2,x_2'=x_2\}$, and proceed
as in Example~\ref{ex:loop-xyz}:
\begin{small}
\[
\begin{array}{|l|l|}
\hline
\multicolumn{1}{|c|}{\transitions_i}  & \multicolumn{1}{c|}{\mbox{Generators of}~\nonegfunc{\proj{\vec{x}}{\transitions_i}}} \\
\hline \hline
\transitions_0{=}\transitions  & \{\mathbf{(1,-1,-1)},(0,0,1)\}\\
\hline
\transitions_1{=}  \transitions_0 \wedge (x_1-x_2)-(x_1'-x_2') \le 0 & \{\mathbf{(-2,1,0)},(1,-1,-1), (0,0,1)\}\\
\hline
\transitions_2{=}  \transitions_1 \wedge (-2x_1+x_2)-(-2x_1'+x_2')\le 0 & \{\mathbf{(2,-1,0)},\mathbf{(-1,0,-1)},(-2,1,0),(0,0,1)\}\\
\hline
\transitions_3{=} \transitions_2 \wedge (2x_1-x_2)-(2x_1'-x_2')\le 0 \wedge & \\
  \hspace*{3.25cm}(-x_1)-(-x_1')\le 0 &\\
\hline
\end{array}
\]
\end{small}%
Let us explain the above steps:
\begin{itemize}
\item $\transitions_0$ is not empty, so we compute the generators of
  $\nonegfunc{\proj{\vec{x}}{\transitions_0}}$, which define the
  nonnegative functions $\rho_1(x_1,x_2,x_3)=x_1-x_2-1$ and
  $\rho_2(x_1,x_2,x_3)=1$, and then compute
  $\transitions_1 = \transitions_0 \wedge \diff{\rho_1}(\vec{x}'')\le
  0 \wedge \diff{\rho_2}(\vec{x}'')\le 0$; and since it differs from
  $\transitions_0$ we recursively call
  $\procdecidemlrf(\transitions_1)$.
\item $\transitions_1$ is not empty, so we compute the generators of
  $\nonegfunc{\proj{\vec{x}}{\transitions_1}}$, which define the
  nonnegative function $\rho_3(x_1,x_2,x_3)=-2x_1+x_2$, and then
  compute
  $\transitions_2 = \transitions_1 \wedge \diff{\rho_3}(\vec{x}'')\le 0$;
    and since it differs from $\transitions_1$ we invoke
    $\procdecidemlrf(\transitions_2)$.
  \item $\transitions_2$ is not empty, so we compute the generators of
    $\nonegfunc{\proj{\vec{x}}{\transitions_2}}$, which define the
    nonnegative functions $\rho_4(x_1,x_2,x_3)=2x_1-x_2$ and
    $\rho_5(x_1,x_2,x_3)=-x_1-1$, and then compute
    $\transitions_3 = \transitions_2 \wedge \diff{\rho_4}(\vec{x}'')\le 0
    \wedge \diff{\rho_5}(\vec{x}'') \le 0$; and since it is equal to
    $\transitions_2$ we return $\transitions_2$.
\end{itemize}
Thus, $\transitions_2$ is a recurrent set of transitions and we
conclude that Loop~\eqref{eq:loop:tiwari} is not
terminating. Projecting $\transitions_2$ on $x_1$ and $x_2$ we get
$\{x_1 \le -1, 2x_1-x_2=0\}$, which is the corresponding recurrent set
of states.

We remark that Loop~\eqref{eq:loop:tiwari} has a fixed point $(-1,-2)$, i.e.,
from state $x_1=-1,x_2=-2$ we have a transition to $x_1=-1,x_2=-2$. 
The algorithm also detects nontermination of some loops that do
not have fixed points. For example, if we change $x_2'=x_2$ in
Loop~\eqref{eq:loop:tiwari} by $x_2'=x_2-1$, we get obtain a
recurrent set of transitions $\poly{S}$ such that
$\proj{\vec{x}}{\poly{S}}=\{-2x_2 \ge 3, 4x_1 -2x_2=1\}$.
\eexm

Now that we have seen the possible outcomes of the algorithm (in case
it terminates), we see that this approach reveals a relation between
seeking \mlrfs and seeking recurrent sets. A possible view is that the algorithm seeks a
recurrent set (of a particular form) and when it concludes that no
such set exists, i.e., reaching $\emptyset$, we can construct
a \mlrf.

The recurrent sets inferred by Algorithm~\ref{alg:mlrf} belong to a narrower class
than that of Definition~\ref{def:recset}.
In fact, for a polyhedral set $\poly{S}$ to be a recurrent set,
Definition~\ref{def:recset} requires that
$\proj{\vec{x}'}{\poly{S}} \subseteq \proj{\vec{x}}{\poly{S}}$, i.e.,
any $\rho(\vec{x}) \ge 0$ that is satisfied by
$\proj{\vec{x}}{\poly{S}}$ is also satisfied by
$\proj{\vec{x}'}{\poly{S}}$.
On the other hand, in our recurrent sets, $\rho$ is monotonic
as well, i.e., 
$\rho(\vec{x}') \ge \rho(\vec{x})$ for any $(\vec{x},\vec{x}')\in\poly{S}$.

\bexm
\label{ex:rsvsmrs}
Consider the following \slc loop:
\begin{equation}
\label{eq:loop:rsvsmrs}
\verb/while / (x \ge 0) \verb/ do / x'= 1-x
\end{equation}
The largest recurrent set of transitions for this loop is
$\{x \ge 0, x \le 1, x'=1-x\}$, and Algorithm~\ref{alg:mlrf} infers
$\{x=\frac{1}{2}, x'=\frac{1}{2}\}$. In the first iteration it
eliminates all transitions for which $x-x'>0$, i.e., those for which
$x\in(\frac{1}{2},\infty)$, and in the second those for which
$(-x)-(-x')>0$, i.e., those for which $x\in[0,\frac{1}{2})$. Note that
this is the largest \emph{monotonic} recurrent set.
\eexm

At this point, it is natural to explore in the difference between the
different notions of recurrent sets. The most intriguing question is
if nonterminating \slc loops always have monotonic recurrent sets
(either polyhedral or closed convex in general). This is clearly true
for loops that have a fixed point, i.e., there is $\vec{x}$ such that
$(\vec{x},\vec{x})\in\transitions$, however, this question is left
open for the general case.
We note that the \emph{geometric nontermination argument} introduced
in~\cite{LeikeH18} is also related to monotonic recurrent sets.
Specifically, it is easy to show that in some cases (when the
coefficients $\mu_i$ and $\lambda_i$, in Definition~5
of~\cite{LeikeH18}, are either $0$ or at least $1$), we can
construct a monotonic recurrent set from the geometric nontermination
argument.

Let us discuss now the case of integer loops. First we note that the
difference between the different notions of recurrent sets is clear in
this case. Example~\ref{ex:rsvsmrs} shows that the loop has a
recurrent set $\{(0,1),(1,0)\}$, but does not have a monotonic
recurrent set.
Apart from this difference, a natural question to ask is if the
recurrent set $\poly{S}$ returned by $\procdecidemlrf(\transitions)$,
or more precisely $\intpoly{\poly{S}}$ if it is not empty, witnesses
nontermination of $\intpoly{\transitions}$. There are some cases for
which this is true.

\blem
Let $\transitions$ be a \slc loop with affine update
$\vec{x}'=U\vec{x}+\vec{c}$ as in~\eqref{eq:affine-update}, and assume
the coefficients $U$ and $\vec{c}$ are integer. If $\poly{S}$ is a
recurrent set of $\transitions$, and $\intpoly{\poly{S}}$ is not
empty, then $\intpoly{\poly{S}}$ is recurrent for
$\intpoly{\transitions}$.
\elem

\bprf
Since the update is affine with integer coefficients, it follows that
any state in $\proj{\vec{x}}{\intpoly{\poly{S}}}$ has a successor in
$\proj{\vec{x}'}{\intpoly{\poly{S}}} \subseteq
\proj{\vec{x}}{\intpoly{\poly{S}}}$, which is the definition of a
recurrent set.
\eprf

We note that one can allow some degree of non-determinism in the above
definition, in particular, nondeterministically setting a variable to
an arbitrary value.
Note that tools for proving nontermination that are based on the use
of Farkas' lemma~\cite{GuptaHMRX08,LarrazNORR14}, impose similar
restrictions to guarantee that the recurrent set is valid over the
integers.
The next example demonstrates that the above lemma does not apply for
\slc loops in general, even when the algorithm is applied to the
integer hull $\inthull{\transitions}$.
This is because it is not guaranteed that any integer state
$\vec{x}\in\intpoly{\proj{\vec{x}}{\poly{S}}}$ has an integer
successor $\vec{x}'\in\intpoly{\proj{\vec{x}'}{\poly{S}}}$.

\bexm
Consider the following loop
\begin{equation}
\verb/while / (x \ge 2 ) \verb/ do / x'=\frac{3}{2}x
\end{equation}
which is clearly nonterminating over the rationals, for any
$x\ge2$. However, it is terminating over the integers because
\begin{inparaenum}[\upshape(\itshape i\upshape)]
\item starting from $x$ odd, the next state $\frac{3}{2}x$ is not integer,
  and;
\item starting from $x$ even, then for some $i>0$ we have
  $\frac{3^i}{2^i}x$ odd (because $\frac{x}{2^i}$ must be odd some
  $i>0$ and $3^i$ is odd), and then the next state in not integer.
\end{inparaenum}
The algorithm returns $\transitions$ as a recurrent set, but
$\intpoly{\transitions}$, which is not empty, is not a recurrent set
as the loop is terminating over the integers.
Note that the transition polyhedron is integral, i.e.,
$\transitions=\inthull{\transitions}$.
\eexm

\subsection{Cases in which Algorithm~\ref{alg:mlrf} does not terminate}
\label{sec:alg:limit}

When Algorithm~\ref{alg:mlrf} terminates, it either finds a \mlrf or
proves nontermination of the given loop. This means that if
applied to a terminating loop that does not have a \mlrf, then
Algorithm~\ref{alg:mlrf} does not terminate.
This nonterminating behaviour shows that our algorithm is not
complete, however, it can also be used to prove that a given loop does
not have a \mlrf.

\bexm
Consider the following loop
\begin{equation}
\label{eq:loop:jan}
\verb/while / (x_1\ge x_2, x_2\ge 1 ) \verb/ do / x_1'=2x_1,\, x_2'=3x_2
\end{equation}
which is terminating~\cite{LeikeHeizmann15}.
It is easy to show (by induction) that the polyhedron passed to
\procdecidemlrf in the $i$-th call is
$\transitions_i=\{ x_1\ge 2^{i}x_2,\,x_2\ge 1,\,x_1'=2x_1,\,
x_2'=3x_2\}$, which satisfies $\transitions_i\neq\emptyset$ and
$\transitions_{i+1}\subset \transitions_{i}$.
This implies that the algorithm does not terminate, and thus the loop
does not have a \mlrf.
Note that $\nonegfunc{\proj{\vec{x}}{\transitions_i}}$ is generated by
the rays $(0,1)$ and $(1,-2^{i})$.
\eexm

The following example shows that Algorithm~\ref{alg:mlrf} might not
terminate also when applied to nonterminating loops.

\bexm
Consider the following loop
\begin{equation}
\label{eq:loop:jan2}
\verb/while / (x_1+x_2 \ge 3) \verb/ do / x_1'=3x_1-2,\, x_2'=2x_2
\end{equation}
which is nonterminating (even over
integers)~\cite{LeikeH18}. Algorithm~\ref{alg:mlrf} does not terminate
on this loop. The loop does have a monotonic recurrent set, e.g.,
$\poly{S}=\{x_1\ge1$, $x_2'=2x_2,\,x_1'=3x_1-2\}$.
\eexm

When the algorithm does not terminate, the iterates $\bkwopiter{\transitions}{i}$ converge to 
$\transitions_\omega=\cap_{i\ge 0}\bkwopiter{\transitions}{i}$.
For example, for Loop~\eqref{eq:loop:jan}, which is terminating, we
have $\transitions_\omega=\emptyset$, and for
Loop~\eqref{eq:loop:jan2}, which is not terminating, we have
$\transitions_\omega=\{x_1\ge1,x_2'=2x_2,\,x_1'=3x_1-2\}$ which is a monotonic
recurrent set.
Some natural questions arise at this point:
\begin{inparaenum}[\upshape(\itshape i\upshape)]
\item is it true that
  $\transitions_\omega=\emptyset$ iff $\transitions$ is terminating?
\item is it a true that if $\transitions_\omega\neq\emptyset$ then it
is a (monotonic) recurrent set?
\end{inparaenum}
We note that for \emph{deterministic} loops,
termination implies $\transitions_\omega=\emptyset$, and it also  not hard to show that if
$\transitions_\omega\neq\emptyset$ then $\transitions_\omega$ is a
monotonic recurrent set.
The general questions are left open.

Once we start to explore properties of
$\transitions_\omega$, there is a difference between real and rational
loops as we demonstrate in the next example.

\bexm
Consider the following loop
\begin{equation}
\label{eq:bravloop}
\verb/while / (4x+y\ge 1 ) \verb/ do / x'=-2x+4y,\, y'=4x
\end{equation}
which is terminating over the rationals and nonterminating over the
reals~\cite{Braverman06}.
The algorithm does not terminate when applied to this loop. If each
$\transitions_i$ is considered as a set of rational transitions, then
$\transitions_w=\emptyset$, however, if we include also the irrational
transitions then $\transitions_w$ would be the closed convex set
$ \{ c \cdot (\sqrt{17}-1, 4, (\sqrt{17}-1)^2, 4(\sqrt{17}-1)) \mid c
\ge \frac{1}{\sqrt{17}+3}\}$, which is a monotonic recurrent set.
\eexm

\section{Loops for which \mlrfs are sufficient}
\label{sec:completeness}

The purpose of this section is to demonstrate the usefulness of
Algorithm~\ref{alg:mlrf} for studying properties of \slc loops. In
particular, we use it characterize kinds of \slc loops for which there
is always a \mlrf, if the loop is terminating.
We shall prove this result for two kinds of loops, both considered in previous work, namely \emph{octagonal relations}
and \emph{affine relations with the finite-monoid property} -- for
both classes, termination has been proven decidable
in~\cite{BIKtacas2012jv}. We only consider the rational case.
Another question, which we do not answer, is whether we can ensure
that Algorithm~\ref{alg:mlrf} recognizes the non-terminating members
of the class.

Let us set some notation first. The composition of transition
relations $S,T \subseteq \rats^{2n}$ is defined as
$S\circ T = \{ \trcv{\vec{x}}{\vec{z}} \mid \exists \vec{y}\ . \
\trcv{\vec{x}}{\vec{y}}\in S \land \trcv{\vec{y}}{\vec{z}}\in T\}$. We
let $T^n=T^{n-1}\circ T$ where $T^0$ is the identity relation. We use
$\pre{n}{T}$ for the projection of $T^n$ over the first $n$
coordinates, i.e., $\pre{n}{T} = \proj{\vec{x}}{T^n}$,
which is the set of states from which we can make traces
of length at least $n$ (for nondeterministic loops some might be less
than $n$ as well).
When $T$ is polyhedral, i.e., a \slc loop, then $T^n$ and
$\pre{n}{T}$ are polyhedral as well.

\subsection{Finite loops}
\label{sec:completeness:finite-loops}

First, we consider loops which always terminate and, moreover, their
number of iterations is bounded by a constant, i.e., there is $N>0$
such that $\transitions^N = \emptyset$.
Note that such loop terminates in at most $N-1$ iterations, or
equivalently $N-1$ is an upper-bound on the length of the
corresponding traces. 

\blem
\label{lem-finite} 
If $\transitions^N=\emptyset$, then it has a \mlrf of depth
less than $N$.
\elem

\bprf
The proof is by induction on $N$.
For $N=1$, $\transitions=\emptyset$, and it has a \mlrf of zero depth,
by definition.
Let $N>1$, and assume that $\transitions^{N-1} \neq \emptyset$,
otherwise it trivially follows for $N$. Consider a transition
$\vec{x}''=(\vec{x},\vec{x}')$ that is the last in a terminating
trace. We have $\vec{x} \in \proj{\vec{x}}{\transitions}$ and
$\vec{x}'\not\in\proj{\vec{x}}{\transitions}$.
Since $\proj{\vec{x}}{\transitions}$ is a closed polyhedral set, this
means that there is a function $\rho$, defined by some
$(\vect{a},b) \in \nonegfunc{\proj{\vec{x}}{\transitions}}$, that is
nonnegative over $\proj{\vec{x}}{\transitions}$ but negative on
$\vec{x}'$, and thus
$\diff{\rho}(\vec{x}'') = \rho(\vec{x})-\rho(\vec{x}') > 0$.
It follows that $\vec{x}''$ is eliminated by Algorithm~\ref{alg:mlrf}
when computing $\transitions'$ at Line~\ref{alg:op:2}.
This means that any transition of $\transitions'$ cannot be the last
transition of any terminating run of $\transitions$, and thus
$(\transitions')^{N-1}=\emptyset$. Therefore, by induction, it has a
\mlrf of depth at most $N-2$, and by Lemma~\ref{lem:mlrf:2}
$\transitions$ has a \mlrf of depth at most $N-1$.
\eprf

 \subsection{The class \dellrf{b}}
 \label{sec:completeness:rfb}

 This class contains loops which can be described as having the
 following behavior: Transitions are linearly ranked, as long as we
 are in states from which we can make runs of length at least $b$. In
 other words, once we reach a state from which we cannot make more
 than $b-1$ transitions we do not require the rest of the trace to
 be linearly ranked. 

 \bdfn
 We say that a \slc loop $\transitions$ belongs to the class
 $\dellrf{b}$ if the loop
 $\transitions \cap \{ (\vec{x},\vec{x}')\in\rats^{2n} \mid
 \vec{x}\in\pre{b}{\transitions} \} $ has a \lrf.
 \edfn
 
 We note that that $\dellrf{1}$ is the class of loops which have a
 \lrf.
 
 \blem
 \label{lem-dellrf}
 Loops in $\dellrf{b}$ have \mlrfs of depth at most $b$.
 \elem

\bprf
This lemma actually generalizes Lemma~\ref{lem-finite}, since the
loops concerned there are $\dellrf{N-1}$. The proof is done
similarly by induction on $b$.
For $b=1$, $\transitions$ has a \lrf by definition.
Let $b>1$, and  suppose that
$\vec{x}''=(\vec{x},\vec{x}') \in \transitions$ is a last transition
of a terminating run, then $\vec{x} \in \proj{\vec{x}}{\transitions}$
and $\vec{x}'\not\in\proj{\vec{x}}{\transitions}$.
Since $\proj{\vec{x}}{\transitions}$ is a closed polyhedral set, this
means that there is a function $\rho$, defined by some
$(\vect{a},b) \in \nonegfunc{\proj{\vec{x}}{\transitions}}$, that is
nonnegative over $\proj{\vec{x}}{\transitions}$ but negative over
$\vec{x}'$, and thus
$\diff{\rho}(\vec{x}'')=\rho(\vec{x})-\rho(\vec{x}') > 0$.
It follows that $\vec{x}''$ is eliminated by Algorithm~\ref{alg:mlrf}
when computing $\transitions'$ at Line~\ref{alg:op:2}.
This means that any transition of $\transitions'$ cannot be the last
transition of any terminating run of $\transitions$, and thus
$\transitions'$ is $\dellrf{b-1}$. Therefore, by induction, it has a
\mlrf of depth at most $b-1$, and by Lemma~\ref{lem:mlrf:2}
$\transitions$ has a \mlrf of depth at most $b$.
\eprf

\bexm
Consider the  loop  (taken from~\cite{BIKtacas2012jv}) defined by
$\transitions=\{ x_2-x_1' \le -1$, \\ $x_3-x_2' \le 0,x_1-x_3' \le
0,x_4'-x_4 \le 0,x_3'-x_4 \le 0\}$.
This loop is \dellrf{3}, since adding
$\pre{3}{Q}=\{x_2+x_4 \ge 1,x_3+x_4 \ge 1,x_1+x_4 \ge 0\}$ to the loop
we find a \lrf, e.g., $\rho(\vec{x})=-x_1-x_2-x_3+3x_4+1$.
Indeed, $\transitions$ has a \mlrf of optimal depth $3$, e.g.,
$\tuple{ -x_1-x_2-x_3+3x_4+1 , -\frac{2}{3}x_1-\frac{1}{3}x_2+x_4+1 ,
  -\frac{1}{4}x_1+\frac{1}{4}x_4+1}$.
Note that the first component is the \lrf  that we have found
above for $\transitions\cap\{\vec{x}'' \mid \vec{x}\in\pre{3}{Q} \}$.
\eexm

Note that if we know that a given class of loop belongs to  $\dellrf{b}$, then
bounding the recursion depth of Algorithm~\ref{alg:mlrf} by $b$ gives
us a decision procedure for the existence of \mlrf for this class.
Bozga, Iosif and Konecn{\'y}~\cite{BIKtacas2012jv} prove that \emph{octagonal relations} are
$\dellrf{5^{2n}}$, where $n$ is the number of variables%
\footnote{Technically, they
prove it just for \emph{integer} loops, but the result 
applies to the rational case as well (one only has to \emph{simplify} some
considerations away from the proof).}.
Thus for octagonal relations, we can decide termination and for terminating loops obtain \mlrf.
For the depth of the \mlrf, namely the parameter $b$ above, \cite{BIKtacas2012jv} gives a tighter (polynomial) result for
those octagonal relations which allow arbitrarily long executions
(called $*$-consistent).

\subsection{Loops with affine-linear updates}
\label{sec:completeness:affine}

In certain cases, we can handle loops with affine-linear updates --
which are, in general, not octagonal.  Recall that a loop with
affine-linear update has a transition relation of the form:
\begin{equation}
  \label{eq:affine-update-loop}
  \transitions \equiv [B\vec{x}\le \vec{b} \land \vec{x}'=U\vec{x}+\vec{c}] \,.
\end{equation}
We keep the meaning of the symbols $U,B,\vec{b},\vec{c}$ fixed for the
sequel. Moreover, we express the loop using the transformation
$\mapsym(\vec{x}) = U\vec{x}+\vec{c}$ and the guard
$\poly{G} \equiv [B\vec{x}\le \vec{b}]$.
We use $U_{ij}$ to denote the entry of matrix $U$ in row $i$ and column
$j$, and for a vector $\vec{v}$ we let $\vec{v}[i..j]$ be the vector
obtained from components $i$ to $j$ of the vector $\vec{v}$.

Our goal is to show that if $U^p$, for some $p>0$, is diagonalizable
and all its eigenvalues are in $\{0,1\}$, then $\transitions$ is
$\dellrf{3p}$, and thus, by Lemma~\ref{lem-dellrf}, if terminating, it
has a \mlrf.
Affine loops with the \emph{finite monoid property} that has been
addressed in~\cite{BIKtacas2012jv}, satisfy this condition (interestingly, in Section~\ref{sec:dp} we show a similar result
when $U-I$ has the property).
We state some auxiliary lemmas first.
 
\blem
\label{lem-squeezelrf}
Let $\transitions$ be an affine-linear loop as
in~\eqref{eq:affine-update-loop} such that, for some $N>0$,
$\transitions^N$ is \dellrf{b}. Then $\transitions$ is
$\dellrf{N(b+1)}$.
\elem

\bprf
If $\transitions^N$ is \dellrf{b}, then
$\transitions^N \cap \{ \vec{x}'' \mid
\vec{x}\in\pre{b}{\transitions^N}\}$ has a \lrf $\rho$, and thus
\begin{equation}
\vec{x} \in \pre{b}{\transitions^N} = \pre{Nb}{\transitions} \
  \Rightarrow\ \rho(\vec x)\ge 0 \,\land\, \rho(\vec{x})-\rho(\mapsym^N(\vec{x})) > 0
  \,.
\end{equation}
Note that $\rho(\vec{x})-\rho(\mapsym^N(\vec{x}))$ can be written as 
\begin{equation}
\label{eq:QN-diff}
  \sum_{j=0}^{N-1} \rho(\mapsym^{j}(\vec{x})) -  \sum_{j=0}^{N-1}\rho(\mapsym^{j+1}(\vec{x}))
\end{equation}
This is because every term $\rho(\mapsym^{i}(\vec{x}))$, except for $i=0$ and
$i=N$, appear in~\eqref{eq:QN-diff} with positive and negative signs.
Hence, if we let $\rho_1(\vec{x}) = \sum_{j=0}^{N-1} \rho(\mapsym^{j}(\vec{x}))$ then:
\[
  \vec{x} \in \pre{Nb}{\transitions} \ \Rightarrow\
  \rho_1(\vec{x})-\rho_1(\mapsym(\vec{x})) > 0 \,.
\]
Moreover, $\rho_1$ is the sum of terms $\rho(\mapsym^{i}(\vec{x}))$ which are
bounded from below on $\pre{N(b+1)}{\transitions}$. Hence, we have a
\lrf for
$\transitions\cap \{ \vec{x}'' \mid
\vec{x}\in\pre{N(b+1)}{\transitions}\}$ and thus $\transitions$ is
$\dellrf{N(b+1)}$.
\eprf

\blem
\label{lem:diag-loop}
Let $\transitions$ be a loop as in~\eqref{eq:affine-update-loop}, and
assume $U$ is diagonal with entries in $\{0,1\}$. Then, if $\transitions$
is terminating, it is $\dellrf{2}$.
\elem

\bprf
Without loss of generality we may assume that
$U_{11} = \dots = U_{kk} = 1$ and $U_{jj} = 0$ for $j>k$, otherwise we
could reorder the variables to put it into this form.
Clearly, the update adds $\vec{c}_1=\vec{c}[1..k]$ to the first $k$
elements of $\vec{x}$, and sets the rest to
$\vec{c}_2=\vec{c}[k+1 .. n]$.
Consequently, such a loop is non-terminating iff the space
$V = \{\vec{x} \in \rats^{n} \mid \vec{x}[k+1 .. n] = \vec{c}_2\}$
intersects the loop guard $\poly{G}\equiv[B\vec{x}\le \vec{b}]$, and
the vector $\vec{u} = (c_1, \dots c_k, 0, \dots, 0)^\trans$ is a
recession direction of the guard, i.e., $B \vec{u} \le \vec{0}$.
To see this: suppose these conditions hold, then starting from any
state $\vec{x}_0 \in V$, the state after $i$ iterations will be
$\vec{x}_i=\vec{x}_0+i\vec{u}$, which is in $\poly{G}$ since
$\vec{x}_0\in\poly{G}$ and $\vec{u}\in\ccone(\poly{G})$, and thus the
execution does not terminate;
for the other direction, suppose it does not terminate, then there
must be a nonterminating execution that starts in $\vec{x}_0\in V$,
this execution generates the states $\vec{x}_0+i\vec{u} \in \poly{G}$
and thus $\vec{u}$ is a recession directions of $\poly{G}$.
 
Now suppose the loop is terminating, we show that it is $\dellrf{2}$.
Let us analyze a run of the loop starting with some valid transition
$(\vec{x}_0, \vec{x}_1)$. We have two cases:

\begin{enumerate}

\item If $\vec{x}_1\not\in \poly{G}$, then the run terminates in $1$
  iteration.

\item If $\vec{x}_1\in\poly{G}$, then $V$ intersects with $\poly{G}$,
  since $\vec{x}_1[k+1..n]=\vec{c}_2$, and thus $B\vec{u} \le \vec{0}$
  should not hold, otherwise the loop is nonterminating. This means
  that there is a constraint $\vect{b}_i\cdot\vec{x}\le b_i$ of the
  guard such that $\vect{b}_i\cdot\vec{u}>0$. Let $\vect{a}$ be as
  $\vect{b}_i$ but setting components $k+1..n$ to zero, we still have
  $\vect{a} \cdot\vec{u}>0$ because these components are $0$ in
  $\vec{u}$.
  We show that this trace is linearly ranked by 
  $\rho(\vec{x})=\vect{a}\cdot\vec{x}+\max(b_i,0)$.
  Suppose the initial state is $\vec{x}_0$, and write it as
  $\trcv{\vec{x}_0[1..k]}{\vec{x}_0[k+1..n]}$.
  Consider a trace that starts in $\vec{x}_0$, it is easy to see
  that the $i$-th state, for $i\ge1$, is
  $\vec{x}_i=\trcv{\vec{x}_0[1..k]+i\vec{c_1}}{\vec{c}_2}$.
  Then, we have
  $\rho(\vec{x}_i)-\rho(\vec{x}_{i+1}) = \vect{a}\cdot\vec{u} > 0$,
  moreover $\rho$ is nonnegative on all state except the last of the
  trace (which is not in the guard).
\end{enumerate}
This analysis implies that any terminating trace is either of length
$1$, or has a \lrf $\rho(\vec{x}) = \vect{a}\cdot\vec{x}+\max(b_i,0)$,
and together with the fact the the loop is deterministic we conclude
that it is \dellrf{2}.
\eprf

Now we are in a position for proving our main result of this section.

\blem  \label{lem:finite-monoid}
If $U^p$, for some $p>0$, is diagonalizable and all its eigenvalues
are in $\{0,1\}$, then loop \eqref{eq:affine-update-loop} is either
nonterminating or $\dellrf{3p}$.
\elem
 
\bprf
Recall that the update is $\mapsym(\vec{x}) = U\vec{x}+\vec{c}$, then
$\mapsym^{p}(\vec{x}) = U^p\vec{x} + \vec{v}$, for a vector
$\vec{v} = (I+U+\dots+U^{p-1})\vec{c}$.  Taking into account the
guard,
\begin{equation}
\label{eq:Qp-loop}
\transitions^p \equiv (B\vec{x}\le \vec{b} \land\dots\land B\mapsym^{p-1}(\vec{x})\le \vec{b}) \land \vec{x}' = \mapsym^p(\vec{x}) \,.
\end{equation}
We write this guard concisely with the notation
$\repl{B}{p} \vec{x} \le \repl{\vec{b}}{p}$.
Since, by assumption, $U^p$ is diagonalizable, there is a non-singular
matrix $P$ and a diagonal matrix $D$ such that $P^{-1} U^p P = D$ and
$D$ has only 1's and 0's on the diagonal ($P$ is a change-of-basis
transformation).
We consider a loop $\widehat{\transitions^p}$ which is similar to
$\transitions^p$ but transformed by $P$, that is:
\begin{equation}
  \label{eq:Qp-loop-similar}
 \widehat{\transitions^p} \equiv \repl{BP}{p}\vec{x}\le \repl{\vec{b}}{p} \land \vec{x}'=D\vec{x}+P^{-1}\vec{v} \,.
\end{equation}
Properties like termination and linear ranking are not affected by
such a change of basis.

This is because if $(\vec{x}_0,\vec{x}_1)$ is a transition of
$\transitions^p$ then $(P^{-1}\vec{x}_0,P^{-1}\vec{x}_1)$ is a
transition of $\widehat{\transitions^p}$, and if
$(\vec{x}_0,\vec{x}_1)$ is a transition of $\widehat{\transitions^p}$
then $(P\vec{x}_0,P\vec{x}_1)$ is a transition of
$\transitions^p$. This means that there is a one-to-one correspondence
between the traces.
Moreover, if function $\vect{a}\cdot\vec{x}+b$ ranks a transition of
$\transitions^p$ then $(\vect{a} P^{-1})\cdot\vec{x}+b$ ranks the
corresponding transition of $\widehat{\transitions^p}$, and if it
ranks a transition $\widehat{\transitions^p}$ then
$(\vect{a}P)\cdot\vec{x}+b$ ranks the corresponding transition of
$\transitions^p$. We conclude that, if terminating, $\transitions^p$
is $\dellrf{b}$ iff $\widehat{\transitions^p}$ is $\dellrf{b}$.

Now, $\widehat{\transitions^p}$ has the diagonal form discussed in
Lemma~\ref{lem:diag-loop}, and thus, in the case that it terminates,
it is $\dellrf{2}$ and so is $\transitions^p$. Then using
Lemma~\ref{lem-squeezelrf} we conclude that $\transitions$ is
$\dellrf{3p}$.
\eprf

\section{\mlrfs and the displacement polyhedron}
\label{sec:dp}

In this section we introduce an alternative representation for \slc
loops, that we refer to as the \emph{displacement polyhedron},
and show that Algorithm~\ref{alg:mlrf}, or more precisely the check
$\bkwopiter{\transitions}{k}=\emptyset$, has a simple encoding in this
representation that can be preformed in polynomial time.
Note that we already know that deciding the existence of a \mlrf of
depth $d$ can be done in polynomial time~\cite{Ben-AmramG17}, so in
this sense we do not provide any new knowledge.
However, the construction is interesting by itself, and, apart from
allowing the efficient encoding of the check
$\bkwopiter{\transitions}{k}=\emptyset$, it provides us with a new approach
to the problem of existence of \mlrfs in general (without
a depth bound), that is still open, and other problems related to termination and nontermination of
\slc loops.
As an indication of their utility, we show that
some nontrivial observations about \slc loops that
are made straightforward through this representation (see the end of
this section).

\bdfn
\label{def:dispol}
Given a \slc loop $\transitions\subseteq\rats^{2n}$, we define its
corresponding \emph{displacement polyhedron} as
$\distransitions =
\proj{\vec{x},\vec{y}}{\transitions\wedge\vec{x}'=\vec{x}+\vec{y}}
\subseteq\rats^{2n}$.
\edfn

Note that the projection drops $\vec{x}'$.  Intuitively, an execution
step using $\transitions$ starts from a state $\vec{x}$, and chooses a
state $\vec{x}'$ such that $\tr{\vec{x}}{\vec{x}'}\in \transitions$.
To perform the step using $\distransitions$, select $\vec{y}$ such
that $\tr{\vec{x}}{\vec{y}}\in\distransitions$ and let the new state
be $\vec{x}+\vec{y}$.  By definition, we obtain the same
transitions. 
The constraint representation of $\distransitions$ can be derived from
that of $\transitions$ as follows.
Let $\transitions\equiv[A''\trcv{\vec{x}}{\vec{x}'}\le \vec{c}'']$ where
$A''$ is the matrix below on the left (see
Section~\ref{sec:prelim:slcloops}), then
$\distransitions\equiv[\dpm\trcv{\vec{x}}{\vec{y}}\le\vec{c}'']$ where
$\dpm$ is the matrix below on the right:
\begin{equation}
\label{eq:dpm}
A'' = \begin{pmatrix} B & 0 \\ A & A' \end{pmatrix}  ~~~~~~~~~\dpm = \begin{pmatrix} B & 0 \\ A+A' & A' \end{pmatrix}
\end{equation}

Next, we show that the displacement polyhedron
$\distransitions_k$ of $\transitions_k=\bkwopiter{\transitions}{k}$ is
equivalent to the following polyhedron projected on $\vec{x}$ and
$\vec{y}_0$
\begin{equation}
  \label{eq:dp:goal}
  \dpm\trcv{\vec{x}}{\vec{y}_0} \le \vec{c}'' \ \land \ \dpm\trcv{\vec{y}_0}{\vec{y}_1} \le \vec{0} \ \land \ \dots \ \land\ \dpm\trcv{\vec{y}_{k-1}}{\vec{y}_{k}} \le \vec{0}
\end{equation}
Now since, by Definition~\ref{def:dispol}, $\transitions_k$ is empty iff
$\distransitions_k$ is empty, the check
$\bkwopiter{\transitions}{k}=\emptyset$ is reduced to checking
that~\eqref{eq:dp:goal} is empty, which can be done in polynomial time
in the bit-size of the constraint representation of $\transitions$ and
the parameter $k$.  We first show how $\distransitions_{k+1}$ can be
obtained from $\distransitions_{k}$ similarly to
$\transitions_{k+1}=\bkwop{\transitions_k}$.

\blem
\label{lem:Rk}
Let $(\vect{a}_1,b_1),\ldots,(\vect{a}_l,b_l)$ are the generators of
the cone $\nonegfunc{\proj{\vec{x}}{\distransitions}}$,
then 
$\distransitions_{k+1} = \distransitions_k \wedge
-\vect{a}_1\cdot\vec{y}\le0\wedge\cdots\wedge-\vect{a}_l\cdot\vec{y}\le0$.
\elem

\bprf
Follows from the fact that
$\proj{\vec{x}}{\transitions_k}=\proj{\vec{x}}{\distransitions_k}$,
and thus $\nonegfunc{\proj{\vec{x}}{\transitions_k}}$ and
$\nonegfunc{\proj{\vec{x}}{\distransitions_k}}$ are the same, and that
for $\rho(\vec x) = \vect{a}\cdot\vec{x} + b$ we have
$\diff{\rho}(\vec{x}'') = \rho(\vec{x})-\rho(\vec{x}') =
-\vect{a}\cdot\vec{y}$, by definition of the displacement polyhedron.
\eprf

\blem
\label{lem:rows-of-B}
Let $(\vect{a}_1,b_1),\ldots,(\vect{a}_l,b_l)$ be the generators of
the cone $\nonegfunc{\proj{\vec{x}}{\distransitions}}$, then
$-\vect{a}_1\cdot\vec{y}\le 0 \wedge\cdots\wedge
-\vect{a}_l\cdot\vec{y}\le 0$ of Lemma~\ref{lem:Rk} is equivalent to $M\vec{y}\le 0$, where
$M$ is such that
$ \proj{\vec{x}}{\distransitions} \equiv [M\vec{x}\le \vec{b}]$.
\elem

\bprf
Consider $(\vect{a},b) \in
\nonegfunc{\proj{\vec{x}}{\transitions}}=\nonegfunc{\proj{\vec{x}}{\distransitions}}$.
By Farkas' lemma, a function $f(\vec{x})=\vect{a}\cdot\vec{x}+b$ is
nonnegative over $\proj{\vec{x}}{\distransitions}$ iff there are
nonnegative $\vect{\lambda}=(\lambda_1,\ldots,\lambda_m)$ such that
$\vect{\lambda}\cdot M=-\vect{a}\wedge\vect{\lambda}\cdot\vec{b}\le
b$.
Note that any (nonnegative) values for $\vect{\lambda}$ define
corresponding values for $\vect{a}$ and $b$. Thus the valid values for
$\vect{a}$ are all conic combinations of the rows of $-M$, i.e., this
cone is generated by the rows of $-M$.
Hence
$-\vect{a}_1\cdot\vec{y}\le 0 \wedge\cdots\wedge
-\vect{a}_l\cdot\vec{y}\le 0$ is equivalent to $M\vec{y}\le \vec{0}$.
\eprf

We use the above lemma to show that
$\distransitions_k$ can be represented as
in~\eqref{eq:dp:goal}, without the need to compute $M$ explicitly.
We first note that using lemmas~\ref{lem:Rk} and~\ref{lem:rows-of-B}
we get that $\distransitions_{k+1} = \distransitions_k \cap \dct_k$,
where
\begin{align*}
\dct_k &= \{ \tr{\vec{x}}{\vec{y} }\in\rats^{2n} \mid M\vec{y} \le \cvz \}  \mbox{~~~~~~~~~\highlight{blue!20}{($M$ as in Lemma~\ref{lem:rows-of-B})}}\\
       & = \{ \tr{\vec{x}}{\vec{y}}\in\rats^{2n} \mid \vec{y} \in \ccone(\proj{\vec{x}}{\distransitions_k})\} \,.
\end{align*}

\blem 
\label{lem:genQk}
$\distransitions_k=\proj{\vec{x},\vec{y}_0}{\distransitions'_k}$ where
$\distransitions'_k$ is defined by~\eqref{eq:dp:goal}.
\elem

\bprf 
We use induction on $k$. For $k=0$ the lemma states that
$\distransitions_0$ is equal
$\dpm\trcv{\vec{x}}{\vec{y}_0} \le \vec{c}'' $, which is correct since
by definition $\distransitions_0=\distransitions$.
Assume the lemma holds for $\distransitions_k$, we prove it for
$\distransitions_{k+1} = \distransitions_k \cap \dct_k$.
By the induction hypothesis,
\begin{equation} 
\label{eq:ind:k}
\distransitions_k= \{\ \trcv{\vec{x}}{\vec{y}_0}\in\rats^{2n}  \mid \dpm\trcv{\vec{x}}{\vec{y}_0} \le \vec{c}'' \ \land \ \dpm\trcv{\vec{y}_0}{\vec{y}_1} \le \vec{0} \ \land \ \dots \ \land\ \dpm\trcv{\vec{y}_{k-1}}{\vec{y}_{k}} \le \vec{0}\ \}
\end{equation}
and
\begin{align*}
 \dct_k = & \{ \tr{\vec{x}}{\vec{y}_0}\in\rats^{2n} \mid \vec{y}_0 \in \ccone(\proj{\vec{x}}{\distransitions_k}) \} 
&\mbox{\llap{\highlight{blue!20}{by definition}}} \\
            = & \{ \tr{\vec{x}}{\vec{y}_0}\in\rats^{2n} \mid \vec{y}_0 \in \ccone(\proj{\vec{x}}{\proj{\vec{x},\vec{y}_0}{\distransitions'_k}}) \} 
&\mbox{\llap{\highlight{blue!20}{by IH}}} \\
            = & \{ \tr{\vec{x}}{\vec{y}_0}\in\rats^{2n} \mid \vec{y}_0 \in \ccone(\proj{\vec{x}}{\distransitions'_k}) \}\\
            = & \{ \tr{\vec{x}}{\vec{y}_0}\in\rats^{2n} \mid \vec{y}_0 \in \proj{\vec{x}}{\ccone(\distransitions'_k)} \} 
&\mbox{\llap{\highlight{blue!20}{by Lemma~\ref{lem:proj.cone}}}} \\
            = & \{\trcv{\vec{x}}{\vec{y}_0}\in\rats^{2n} \mid \dpm\trcv{\vec{y}_0}{\vec{y}_1} \le \vec{0} \land  \dpm\trcv{\vec{y}_1}{\vec{y}_2} \le \vec{0} \land \cdots \land  \dpm\trcv{\vec{y}_{k}}{\vec{y}_{k+1}} \le \vec{0} \}
\end{align*}
Note that in the last step, we incorporated the recession cone of $\distransitions_k'$ as in~\eqref{eq:dp:goal}, after
renaming $\vec{y}_i$ to $\vec{y}_{i+1}$, and $\vec{x}$ to
$\vec{y}_0$ just to make it easier to read in the next step.
Now, let us compute $\distransitions_{k+1}=\distransitions_k\cap
\dct_k$. Note that any $\trcv{\vec{x}}{\vec{y}_0} \in
\distransitions_{k+1}$ must satisfy the constraint $\dpm\trcv{\vec{x}}{\vec{y}_0} \le
\vec{c}''$ that comes form $\distransitions_k$. Adding this constraint
to $\dct_k$ above we clearly obtain a subset of
$\distransitions_k$, and thus
\[
 \distransitions_{k+1} = \{\trcv{\vec{x}}{\vec{y}_0}\ \mid \dpm\trcv{\vec{x}}{\vec{y}_0} \le \vec{c}''\ \land\ \dpm\trcv{\vec{y}_0}{\vec{y}_1} \le \vec{0} \land \ \cdots \ \land\ \dpm\trcv{\vec{y}_{k-1}}{\vec{y}_{k}} \le \vec{0} \}
\]
which is exactly $\proj{\vec{x},\vec{y}_0}{\distransitions'_{k+1}}$,
justifying the lemma's statement for $k+1$.
\eprf

\blem
$\transitions$ has a \mlrf of depth $d$ iff
$\distransitions'_{d}$ is empty.
\elem

\bprf
By Lemma~\ref{lem:mlrf} $\transitions$ has a \mlrf of depth $d$ iff
$\transitions_d=\bkwopiter{\transitions}{d}$ is empty, and by
Definition~\ref{def:dispol} $\transitions_d$ is empty iff
$\distransitions_d$ is empty, and since $\distransitions_d$ is empty
iff $\distransitions'_d$ is empty the lemma follows.
\eprf

Apart from providing a polynomial-time implementation for the check
that $\bkwopiter{\transitions}{d}=\emptyset$, we believe that the 
above lemma provides us with a new tool for addressing the
problem of deciding whether a given \slc loop has a $\mlrf$ of any
depth, which is still an open problem. Next we discuss some
directions.

One direction is to come up with conditions on
the matrices $A''$ (or equivalently $\dpm$) and $\vec{c}''$ under
which it is guaranteed that if $\distransitions'_k$ is empty then $k$
must be smaller than some $d$, i.e., bounding the depth of \mlrfs for
classes of loops that satisfy these conditions.
We can also view the problem as looking for some $N$, such that
$\poly{C}^N=\poly{C}^{N+1}$ where
$\poly{C}=[\dpm\trcv{\vec{y}}{\vec{y}'} \le \vec{0}]$, which is a
sufficient condition for the algorithm to terminate in at most $N$
iterations, since then $\distransitions_N=\distransitions_{N+1}$,
either with a recurrent set or with a \mlrf.
This is particularly interesting if the loop is deterministic with
affine update as in~\eqref{eq:affine-update}. In such case
$\poly{C}=[B\vec{y}\le0\wedge\vec{y}'=(U-I)\vec{y}]$, where
$I\in\rats^{n\times n}$ is the identity matrix, and thus if the matrix
$(U-I)$ is nilpotent, for example, then there is $N$ such that
$\poly{C}^N=\poly{C}^{N+1}$. This also holds when matrix $(U-I)$
satisfies the finite-monoid property discussed in
Section~\ref{sec:completeness:affine}.

Another tantalizing observation reduces the existence of $d$ such that
$\distransitions'_d$ is empty to the question whether a related
\slc loop terminates, for a given polyhedron of initial states, in a bounded number of steps.
Specifically,  the loop:
\[
  \verb/while / (B\vec{y} \le \vec{0}) \verb/ do / (A+A')\vec{y} + A'\vec{y}'\le\vec{0}.
\]
where $B$, $A$ and $A'$ are those used in the definition of $\dpm$, and the question whether
it terminates in at most $d$ steps for all $\vec{y}\in\{ \vec{y}\in\rats^n \mid
\dpm\trcv{\vec{x}}{\vec{y}}\le\vec{c}''\}$.
Note that if the update is affine as in~\eqref{eq:affine-update}, then the
above loop is equivalent to
\[
  \verb/while / (B\vec{y} \le \vec{0}) \verb/ do / \vec{y}'=(U-I)\vec{y}
\]
where $I\in\rats^{n\times n}$ is the identity matrix.

To further demonstrate the usefulness of the displacement polyhedra,
next we provide some observations, regarding \slc loops with bounded
transition polyhedra, that are easy to see using the displacement
polyhedron and are much less obvious using the transition polyhedron.
Recall that a transition polyhedron is bounded if its recession cone
consists of a single point $(\vec{0},\vec{0})$.

\blem
A bounded \slc loop $\transitions$ it is nonterminating iff it has a
fixpoint $(\vec{x},\vec{x})\in\transitions$, and it is terminating iff
it has a \lrf.
\elem

\bprf
Since $\transitions$ is bounded, the displacement
polyhedron
$\distransitions\equiv[\dpm\trcv{\vec{x}}{\vec{y}}\le\vec{c}'']$ is
bounded, and its recession cone $\dpm\trcv{\vec{x}}{\vec{y}}\le\vec{0}$
consists of a single point $(\vec{0},\vec{0})$.
From the form of $\distransitions'_k$, which is a conjunction of
instances of $\dpm\trcv{\vec{y}_i}{\vec{y}_{i+1}}\le\vec{0}$, it is
easy to see that $\distransitions_2=\distransitions_1$. This means
that the algorithm will terminate in at most two iterations with one
of the following outcomes:
\begin{inparaenum}[\upshape(\itshape i\upshape)]
\item $\distransitions_0=\distransitions_1$;
\item $\distransitions_2=\distransitions_1$; or
\item $\distransitions_1$ is empty.
\end{inparaenum}  
In the first two cases all transitions of $\distransitions_1$ or
$\distransitions_2$ are of the form $\trcv{\vec{x}}{\vec{0}}$, and thus
$(\vec{x},\vec{x}) \in \transitions$; in the third case we have found
a \mlrf of depth $1$, i.e., \lrf.
Note that the part that relates nontermination to the existence of a
fixpoint follows also from~\cite{LeikeH18} as well.
\eprf

\section{Conclusions}
\label{sec:conc}

The purpose of this work has been to improve our understanding of
\mlrfs, in particular of the problem of deciding whether a given \slc
loop has a \mlrf without a given bound on the depth. The outcomes are
important insights that shed light on the structure of these ranking
functions.

At the heart of our work is an algorithm that seeks \mlrfs, which is
based on iteratively eliminating transitions, that satisfy some
property, until eliminating them all or stabilizing on a set of
transitions that cannot be reduced anymore.
In the first case, a \mlrf can be constructed, and, surprisingly, in
the second case the stable set of transitions turned to be a recurrent
set that witnesses nontermination. This reveals an equivalence between
the problems of seeking \mlrfs and seeking recurrent sets of a
particular form.

Apart from the relation to seeking recurrent sets, the insights of our
work are helpful for characterizing classes of loops for which there
is a always a \mlrf, when terminating. We demonstrated this for two
classes that have been considered previously.
In addition, our insights led to a new representation for \slc in
which our algorithm has a very simple formalization, and, moreover, it
allows making nontrivial observations regards (bounded) \slc loop
straightforward. We believe that this representation can be useful for
other related problems.
Our research leaves as well a considerable amount of related
\emph{new open questions}, which we hope will trigger the interest of
the community in this line of work.

An implementation of Algorithm~\ref{alg:mlrf} is available at
\url{http://www.loopkiller.com/irankfinder} -- options $\mlrf(\rats)$
or $\mlrf(\ints)$ -- together with corresponding examples of \slc
loops.
For experimentally evaluating Algorithm~\ref{alg:mlrf} for
nontermination, we have integrated it in a newer version of
\irankfinder that takes as input a control-flow graph, as follows:
when it fails to prove termination, it enumerates closed walks (which
are basically \slc loops) using only transitions whose termination was
not proven, and then applies Algorithm~\ref{alg:mlrf} to seek
recurrent sets. For now it does not check that the recurrent set is
reachable, which is an orthogonal problem.
We have analyzed $439$ benchmarks taken from \texttt{TPDB}~\cite{tpdb}
for which \irankfinder fails to prove termination -- For $400$ it find
recurrent sets, out of which $273$ include a fixpoint transition. For
the remaining $127$ benchmarks, $114$ are defined by octagonal
relations and $114$ are deterministic.
All are available at \url{http://irankfinder.loopkiller.com}, under
the folder \texttt{Papers/mlrf\_recset}, and can be analyzed by
enabling \texttt{Nontermination} in the settings section.

The problem of seeking \mlrfs with a given bound on the depth has been
considered before in several works. The complexity and algorithmic
aspects has been completely settled in~\cite{Ben-AmramG17}. \mlrfs for
general loops have been considered
in~\cite{LeikeHeizmann15,li2016depth}, where both use non-linear
constraint solving. In~\cite{BagnaraM13PPDP} the notion of ``eventual
linear ranking functions," which are \mlrfs of depth 2 has been
studied. The approach described in~\cite{BorrallerasBLOR17} is also
able to infer \mlrfs for general loops incrementally, by solving
corresponding safety problems using Max-SMT.
Lexicographic ranking function are very related, their algorithmic and
complexity aspects has been considered in several
works~\cite{DBLP:conf/cav/BradleyMS05,ADFG:2010,Ben-AmramG13jv,LarrazORR13}.

Nontermination of programs or corresponding control-flow graphs has been
considered before in several works:
some~\cite{GuptaHMRX08,LarrazNORR14,PayetMS14,BakhirkinP16,BakhirkinBP15,BrockschmidtSOG11,BIKtacas2012jv,LeikeH18}
are based on finding recurrent sets in one form or another; while
others are based on reducing the problem to proving non-reachability
of terminating states~\cite{ChenCFNO14,VelroyenR08}.
 
\bibliographystyle{plain}

\end{document}